\newif\ifAlpha \Alphafalse

\newif \ifhyperlinks    \hyperlinkstrue

\newif \ifDraft \Draftfalse
\newif\ifFinal \Finalfalse

\documentclass[9pt\ifFinal\else,preprint,nocopyrightspace\fi,\ifAlpha\else natbib,authoryear\fi]{sigplanconf}
\ifAlpha\else\bibpunct{[}{]}{;}{a}{}{,}\fi 
\usepackage{amsmath,times}
\usepackage[pdftex]{graphicx}

\usepackage{amsmath,times,amsthm}
\usepackage[pdftex]{graphicx}
\usepackage{float}
\usepackage{mathpartir}
\usepackage{amsmath}
\usepackage{amssymb}
\usepackage{stmaryrd}
\usepackage{txfonts}
\usepackage{ulem}
\usepackage{util}
\usepackage{haskell}
\usepackage{placeins}
\usepackage{xspace}
\usepackage{xfrac}
\usepackage{tikz}
\usepackage{alltt}
\usetikzlibrary{shapes.arrows,chains,positioning}
\usepackage{paralist}           
\usepackage{color}

\definecolor{commentcol}{rgb}{0.3,0.3,0.3}
\definecolor{keywordcol}{rgb}{0,0,0.4}
\definecolor{typecol}{rgb}{0.4,0.1,0}
\definecolor{funccol}{rgb}{0.1,0.4,0}
\definecolor{anticol}{rgb}{0.1,0.4,0.4}
\definecolor{light-gray}{gray}{0.90}
\usepackage{listings}
\lstdefinelanguage{CDSL}{
  basicstyle=\ttfamily\scriptsize,
  morecomment=[l]{--},
  morekeywords={type, let, in, and, if, then, else},
}

  \lstdefinestyle{isa}{
    basicstyle=\ttfamily\scriptsize,  
    columns=fullflexible,
    keepspaces,
    mathescape=true,
    morekeywords={locale,fixes,assumes,shows,and,lemma,definition,in,
      type_synonym,where,procedures,theorem,if,then,case,of,let},
    literate=
      {"}{}0
      {ÔøΩ}{$^\prime$}1
      {'}{$^\prime$}1
      {\\<^sub>6}{$_6$}1      
      {\\<^sub>3}{$_3$}1      
      {\\<^sub>2}{$_2$}1      
      {\\<^sub>1}{$_1$}1      
      {\\<^sub>0}{$_0$}1      
      {\\<^sub>f}{$_f$}1      
      {\\<forall>}{$\forall$}1
      {\\<exists>}{$\exists$}1
      {\\<equiv>}{$\equiv$}1
      {\\<Longrightarrow>}{$\Longrightarrow$}2
      {\\<longrightarrow>}{$\longrightarrow$}2
      {\\<Rightarrow>}{$\Rightarrow$}1
      {\\<rightarrow>}{$\rightarrow$}1
      {\\<longleftrightarrow>}{$\longleftrightarrow$}2
      {\\<Down>}{$\Downarrow$}2
      {\\<guillemotleft>}{$\mathopen{\{\mkern-4mu|}$}2
      {\\<guillemotright>}{$\mathclose{|\mkern-4mu\}}$}2
      {\\<And>}{$\bigwedge$}1
      {\\<and>}{$\land$}1
      {\\<or>}{$\lor$}1
      {\\<not>}{$\lnot$}1
      {\\<in>}{$\in$}1
      {\\<notin>}{$\notin$}1
      {\\<noteq>}{$\neq$}1
      {\\<le>}{$\le$}1
      {\\<ge>}{$\ge$}1
      {\\<cup>}{$\cup$}1
      {\\<cap>}{$\cap$}1
      {\\<lambda>}{$\lambda$}1
      {\\<times>}{$\times$}1
      {\\<turnstile>}{$\vdash$}1
      {\\<turnstile-t>}{$\vdash_t$}2
      {\\<subseteq>}{$\subseteq$}1
      {\\<lbrace>}{$\mathopen{\{\mkern-4mu|}$}1
      {\\<rbrace>}{$\mathclose{|\mkern-4mu\}}$}1
      {\\<lbrakk>}{$\mathopen{\lbrack\mkern-3mu\lbrack}$}1
      {\\<rbrakk>}{$\mathclose{\rbrack\mkern-3mu\rbrack}$}1
      {\\<infinity>}{$\infty$}1
      {\\<sigma>}{$\sigma$}1
      {\\<mu>}{$\mu$}1
      {\\<gamma>}{$\gamma$}1
      {\\<tau>}{$\tau$}1
      {\\<Gamma>}{$\Gamma$}1
       {\\<le>}{$\leq$}1
      {\\<Sum>}{$\sum$}1
      }

  \lstnewenvironment{isalst}{\lstset{style=isa,
    basicstyle=\footnotesize}}{}

\renewcommand{\cite}[1]{\errmessage{Don't use cite, you probably want citep}}
\usepackage{verbatim}           

\usepackage{xspace}             

\usepackage{balance}            

\newcommand{\code}[1]{\texttt{#1}}

\ifDraft
\newcommand{\STATUS}[2]{\fbox{\textsf{\textcolor{orange}{Owner: }}{\textbf{#1}}\quad{\textsf{\textcolor{orange}{Status: }{#2}}}}}
\else
\newcommand{\STATUS}[2]{}{}
\fi

\usepackage[pdftex]{graphicx}
\setkeys{Gin}{keepaspectratio=true,clip=true,draft=false,width=\linewidth}
\graphicspath{{./imgs/}}

\ifDraft
  \usepackage{draftwatermark}
  \SetWatermarkLightness{0.85}
  \newcommand{\Comment}[1]{\textbf{\textsl{#1}}}
  \newenvironment{LongComment}[1] 
    {\begingroup\par\noindent\slshape \textbf{Begin Comment[#1]}\par}
    {\par\noindent\textbf{End Comment}\endgroup\par}
  \newcommand{\FIXME}[1]{\textbf{\textsl{\colorbox{yellow}{FIXME:} #1}}}
  \newcommand{\TODO}[1]{\textbf{\textsl{TODO: #1}}}
\else
  \newcommand{\Comment}[1]{\relax}
  
  \newcommand{\FIXME}[1]{\relax}
  \newcommand{\TODO}[1]{\relax}
\fi

\newcommand{\ignore}[1]{}

\usepackage[pdftex]{hyperref}
\ifhyperlinks 
    \hypersetup{
      colorlinks,
      linkcolor={red},
      citecolor={blue},
      urlcolor={blue} 
    }
\else
  \hypersetup{nolinks=true}
\fi

\begin{document}

\newtheorem{lemma}{Lemma}
\newtheorem{definition}{Definition}
\newtheorem{assumption}{Assumption}
\newtheorem{theorem}{Theorem}

\floatstyle{boxed}
\restylefloat{table}
\restylefloat{figure}

\newcommand{\CDSL}{\textsc{Cogent}\xspace}
\newcommand{\cdsl}{\CDSL}
\newcommand{\DDSL}{DDSL\xspace}
\newcommand{\todo}[1]{\TODO{#1}}

\newcommand{\Cone}{\ding{172}\xspace}
\newcommand{\Ctwo}{\ding{173}\xspace}
\newcommand{\Cthree}{\ding{174}\xspace}
\newcommand{\Cfour}{\ding{175}\xspace}

\normalem

  \def\Snospace~{\S\nobreak\hspace{0.1ex}{}}
  \renewcommand{\figureautorefname}{Fig.}
  \renewcommand{\sectionautorefname}{\Snospace}
  \renewcommand{\subsectionautorefname}{\Snospace}
  \renewcommand{\subsubsectionautorefname}{\Snospace}
  \renewcommand{\appendixautorefname}{Appendix}
  \renewcommand{\Hfootnoteautorefname}{Footnote}
  \newcommand{\Htextbf}[1]{\textbf{\hyperpage{#1}}}

  \makeatletter
  \newsavebox{\@brx}
  \newcommand{\llangle}[1][]{\savebox{\@brx}{\(\m@th{#1\langle}\)}%
    \mathopen{\copy\@brx\mkern2mu\kern-0.9\wd\@brx\usebox{\@brx}}}
  \newcommand{\rrangle}[1][]{\savebox{\@brx}{\(\m@th{#1\rangle}\)}%
    \mathclose{\copy\@brx\mkern2mu\kern-0.9\wd\@brx\usebox{\@brx}}}
  \makeatother

  \title{\CDSL: Certified Compilation for a Functional Systems Language}


\authorinfo{%
Liam~O'Connor,
Christine~Rizkallah,
Zilin~Chen,
Sidney~Amani,
Japheth~Lim,
Yutaka~Nagashima,
Thomas~Sewell,
Alex~Hixon,
Gabriele~Keller,
Toby~Murray,
Gerwin~Klein}
  {NICTA, Sydney, Australia \\
            University of New South Wales, Australia}
  {\href{mailto:christine.rizkallah@nicta.com.au}{first.last@nicta.com.au}%
    \ifDraft Draft of \today \pageref{p:lastpage} pages (of 12 total excl. bib). \fi}

  \maketitle

  \urlstyle{sf}
  \thispagestyle{empty}
  \begin{abstract}

    We present a self-certifying compiler for the \CDSL systems language.
    \CDSL is a restricted, polymorphic, higher-order, and purely functional
    language with linear types and without the need for a trusted runtime or
    garbage collector. It compiles to efficient C code that is designed to
    interoperate with existing C functions.
    The language is suited for layered systems code with minimal sharing such
    as file systems or network protocol control code.
     
    For a well-typed \CDSL program, the compiler produces C code, a
    high-level shallow embedding of its semantics in Isabelle/HOL, and a
    proof that the C code correctly implements this embedding. The aim is
    for proof engineers to reason about the full semantics of
    real-world systems code productively and equationally,
    while retaining the interoperability and leanness of C.
    
    We describe the formal verification stages of the compiler, which include
    automated formal refinement calculi, a switch from imperative update
    semantics to functional value semantics formally justified by the linear
    type system, and a number of standard compiler phases such as type
    checking and monomorphisation. The compiler certificate is a series of
    language-level meta proofs and per-program translation validation phases,
    combined into one coherent top-level theorem in Isabelle/HOL.

  \end{abstract}

 \ifFinal
  \pagestyle{empty}
\fi

\ifFinal
\fi

\ifFinal, domain-specific languages, file systems, Isabelle/HOL\fi

\section{Introduction}\label{s:intro}

Imagine writing low-level systems code in a purely functional language and then
reasoning about this code equationally and productively in an interactive
theorem prover. Imagine doing this without the need for a trusted compiler,
runtime or
garbage collector and letting this code interoperate with native C parts of
the system, including your own efficiently implemented and formally verified
additional data types and operations.

\CDSL achieves this goal by certified compilation from a high-level, pure,
polymorphic, functional language with linear types, specifically designed for
certain classes of systems code. For a given well-typed \CDSL program, the
compiler will produce a high-level shallow embedding of the program's
semantics in Isabelle/HOL~\citep{Nipkow_Klein:Isabelle}, and a theorem that
connects this shallow embedding to the C code that the compiler produces:~any 
property proved of the shallow embedding is guaranteed
to hold for the generated C.

The compilation target is C, because C is the language most existing 
systems code is written in, and because with the advent of tools like
CompCert~\citep{Leroy_06,Leroy_09} and gcc translation
validation~\citep{Sewell_MK_13}, C is now a language with well understood
semantics and existing formal verification infrastructure.

If C is so great, why not verify C systems code directly? After all, there is
an ever growing list of successes
\citep{Klein_EHACDEEKNSTW_09,Klein_AEMSKH_14,Gu_KRSWWZG_15,Beringer_PYA_15}
in this space. The reason is simple: verification of manually written C
programs remains expensive. Just as high-level languages increase
programmer productivity, they should also increase verification productivity.
Certifying compilation of a language with verification-friendly semantics is
a key step in achieving this goal for \CDSL.

The state of the art for certified compilation of a full featured
functional language is
CakeML~\citep{Kumar_MNO_14}, which covers an entire ML dialect. \CDSL is
targeted at a substantially different point in the design space. CakeML includes a verified runtime and garbage collector,
while \CDSL
works hard to avoid these so it can be applicable to low-level embedded
systems code. CakeML covers full turing-complete ML with complex semantics
that works well for code written in theorem provers.
\CDSL is a restricted language of total functions with intentionally simple
semantics that are easy to reason about equationally.
CakeML is great for application code; \CDSL is great for systems code,
especially layered systems code with minimal sharing such as the control
code of file systems or network protocol stacks. \CDSL is not designed for
systems code with closely-coupled, cross-cutting sharing, such as
microkernels.

\CDSL's main restrictions are the (purposeful) lack of recursion and
iteration and its linear type system. The former ensures totality,
which is important for both systems code correctness as well as for a simple
shallow representation in higher-order logic. The latter is important for
memory management and for making the transition from imperative C semantics
to functional value semantics. Even in the restricted target domains of
\CDSL, real programs will of course contain some amount of iteration. 
This is
where \CDSL's integrated foreign function interface comes in: the engineer 
provides her own verified data types and
iterator interfaces in C and uses them seamlessly in \CDSL, including in
formal reasoning.

\CDSL is restricted, but it is not a toy language. We have used it to
implement two efficient full-scale Linux file systems --- a custom Flash file system
and an implementation of standard Linux ext2. We plan to report on the experience with these implementations in
separate work. The focus of this paper is what can be learned 
from \CDSL about the formal verification of certifying compilation.

In particular, this paper discusses in detail the following contributions:
\begin{inparaenum}[a)]
\item the self-certifying \CDSL compiler and language;
\item the formal semantics of the \CDSL language and the switch from
imperative update semantics to functional value semantics formally justified
by the linear type system (\autoref{s:lang});
\item the top-level compiler certificate (\autoref{s:toplevel}), which is a series of language-level
meta proofs and per-program translation validation phases;
\item the verification stages that make up the correctness theorem (\autoref{s:verification}), including
automated refinement calculi, formally verified type checking, A-normalisation, and monomorphisation; and
\item the lessons learned in this project on functional language formalisation and
compiler correctness proofs (\autoref{s:lessons}).
\end{inparaenum}

\section{Overview}\label{s:overview}

Our aim in this paper is to build a self-certifying compiler from \cdsl to
efficient C code, such that a proof engineer can reason equationally about
its semantics in Isabelle/HOL and apply the compiler theorem to derive
properties about the generated C code. Formally, the certificate theorem is a
refinement statement between the shallow embedding and the C code. This
generated C code can be compiled by CompCert. It also falls into the subset
of the gcc translation validation tool by \citet{Sewell_MK_13}, whose theorem
would compose directly with our compiler certificate.\footnote{At the time of writing, \cdsl's occasionally larger stack frames lead to gcc emitting
\texttt{memcpy()} calls that, while conceptually straightforward to handle,
the translation validator does not yet cover.}

Shallow embeddings are nice for the human user, but they do not provide much
syntactic structure for constructing the compiler theorem.
Therefore, the compiler also generates a deep embedding for each \CDSL
program to use in the internal proof chain.
There are two semantics for this deep embedding.
\begin{inparaenum}[(1)]
\item a formal functional \emph{value semantics} where programs evaluate to values and 
\item a formal imperative \emph{update semantics} where programs manipulate references to mutable global state.
\end{inparaenum}

\floatstyle{plain} \restylefloat{figure}
\begin{figure}[tbh]
    \begin{center}
      \includegraphics[width=\columnwidth]{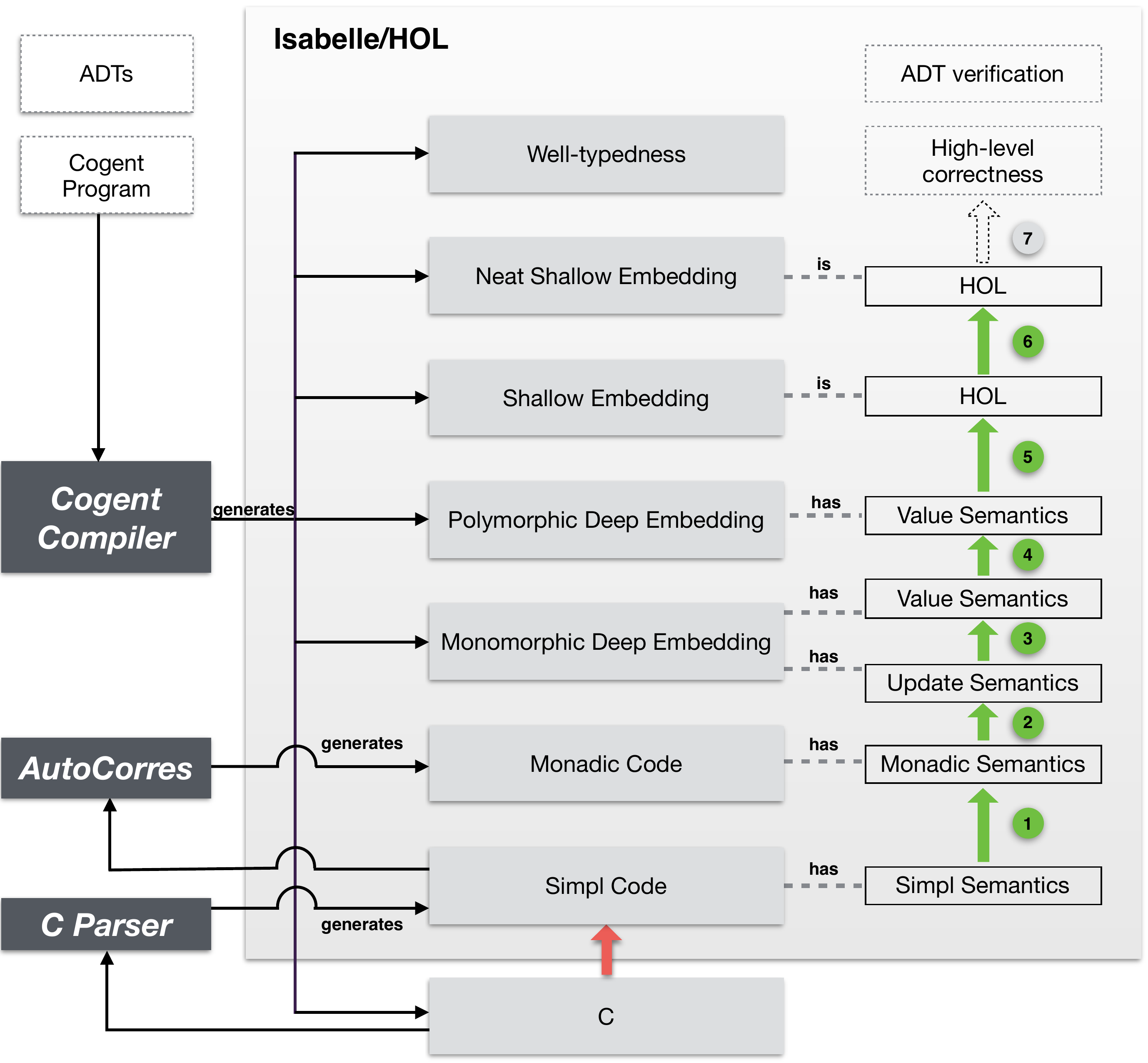}
    \end{center}
    \caption{A detailed overview of the verification chain.}
    \label{fig:refinement}
  \end{figure}
\floatstyle{boxed}
\restylefloat{figure}

\noindent 
\autoref{fig:refinement} shows an overview of the program representations
generated by the compiler and the break-down of the automatic refinement
proof that makes up the compiler certificate. The program representations
are, from the bottom of \autoref{fig:refinement}: the C code, the semantics
of the C code expressed in Isabelle/Simpl~\citep{schirmer:phd}, the same
expressed as a monadic functional
program~\citep{Greenaway_AK_12,Greenaway_LAK_14}, a monomorphic A-normal deep
embedding of the \cdsl program, a polymorphic A-normal deep embedding of the
same, an A-normal shallow embedding, and finally a `neat' shallow embedding
of the \cdsl program that is syntactically close to the \cdsl input of the
compiler. Most of the theorems assume that the \cdsl program is well-typed,
which is discharged automatically in Isabelle with type inference
information from the compiler.

The solid arrows on the right-hand side of the figure represent refinement
proofs and the labels on these arrows correspond to the numbers in the
following description. The only arrow that is not formally verified is the
one crossing from C code into Isabelle/HOL at the bottom of
\autoref{fig:refinement} --- this is the C-to-Isabelle parser~\citep{Tuch_KN_07},
which is a mature verification tool used in a number of large-scale
verifications. As mentioned, it could additionally be checked by translation
validation. We briefly describe each intermediate theorem, starting with the
Simpl code at the bottom of the figure. For well-typed \cdsl programs, we
automatically prove:

\begin{enumerate}
  \item Theorem: The Simpl code produced by the C parser corresponds to a
    monadic representation of the C code. 
    The proof is generated using an adjusted version of the AutoCorres tool.
 \item Theorem: The monadic program terminates and is a refinement of the monomorphic \CDSL deep embedding 
    under the update semantics.
 \item Theorem: If a \CDSL deep embedding evaluates in the update
semantics then it evaluates to the same result in the value semantics.
This is a known consequence of linear type systems~\citep{Hofmann_00}, but to our knowledge
it is the first mechanised proof of such a property, esp.\ for a full-scale language.
 \item Theorem: If a monomorphic \CDSL deep embedding evaluates in the value semantics 
     then the polymorphic deep embedding evaluates equivalently in the value semantics.
 \item Theorem: If the polymorphic \CDSL deep embedding evaluates in the value semantics then 
     the \CDSL shallow embedding evaluates to a corresponding shallow Isabelle/HOL value.
 \item Theorem: The A-normal shallow embedding is (extensionally) equal in
Isabelle/HOL to a syntactically neater shallow embedding, which is more
convenient for human reasoning. This human-friendly shallow embedding
corresponds to the \CDSL code before the compiler's A-normalisation phase.
\end{enumerate}

\noindent
Arrow 7 indicates verification of user-supplied abstract data types
(ADTs) implemented in C and further manual high-level proofs on top of the
human-friendly shallow embedding. These are enabled by the previous steps,
but are not part of this paper.

In \autoref{s:verification} we define in more detail the relations that
formally link the values (and states, when applicable) that these programs
evaluate to. Steps (3) and (4) are general properties about the language and
we therefore prove them manually once and for all. Steps (1), (2), (5), and
(6) are generated by the compiler for every program. The proof for step (1)
is generated by AutoCorres. For steps (2) and (5) we define compositional
refinement calculi that ease the automation of these proofs. Step (6), the
correctness of A-normalisation, is straightforward to prove via rewriting
because at this stage we can already use equational reasoning.

\section{Language}\label{s:lang}

In this section we formally define \CDSL, including its linear type system,
its two dynamic semantics --- update and value --- mentioned earlier in
\autoref{s:overview}, and the refinement theorem between them. We begin the section by walking through an example
\CDSL programs.

\subsection{Example}

\autoref{fig:cdsl-snippet} shows an excerpt of our \CDSL ext2 implementation.
The example uses not all, but many features of the language.




\begin{figure}[th]
\begin{lstlisting}[language=CDSL]
type ExSt
type UArray a
type Opt a = <None () | Some a>
type Node = #{mbuf:Opt Buf, ptr:U32, fr:U32, to:U32}
type Acc = (ExSt, FsSt, VfsInode)
type Cnt = (UArray Node, 
  (U32, Node, Acc, U32, UArray Node) -> (Node, Acc))

uarray_create: all (a :< E). (ExSt, U32) 
  -> <Success (ExSt, UArray a) | Err ExSt>

ext2_free_branch: (U32, Node, Acc, U32) 
  -> (Node, Acc, <Expd Cnt | Iter ()>
ext2_free_branch (depth,nd,(ex,fs,inode),mdep) =
  if depth + 1 < mdep 
    then
      uarray_create[Node] (ex,nd.to-nd.fr) !nd
      | Success (ex, children) =>
        let nd_t { mbuf } = nd
        and (children, (ex, inode, _, mbuf)) = 
          uarray_map_no_break #{
            arr  = children,
            f    = ext2_free_branch_entry,
            acc  = (ex, inode, node_t.fr, mbuf),
            ... } !nd_t
        and nd = nd_t { mbuf }
        in (nd, (ex, fs, inode), 
          Expd (children, ext2_free_branch_cleanup))
      | Err ex -> (nd, (ex,fs,inode), Iter ())
    else ...
\end{lstlisting}
\caption{\CDSL example}\label{fig:cdsl-snippet}
\end{figure}

\newcommand{\cdslfunc}[1]{\texttt{\textcolor{funccol}{#1}()}\xspace}
\newcommand{\cdslvar}[1]{\texttt{#1}\xspace}
\newcommand{\cdsltype}[1]{\texttt{\textcolor{typecol}{#1}}\xspace}
\newcommand{\cdsltypenospace}[1]{\texttt{\textcolor{typecol}{#1}}}

\noindent
The first line in \autoref{fig:cdsl-snippet} shows the \cdsl side of the
foreign function interface. It declares an abstract \CDSL data
type~\cdsltype{ExSt}, implemented in C. Line 2 shows a parametric abstract
type, and line 9 shows a corresponding abstract
function~\cdslfunc{uarray\_create}, also implemented in C. Note that this
abstract function is polymorphic, with a kind constraint~$\Escapable$ (see
\autoref{s:kinding}) on type argument \cdslvar{a}.

The integration of such foreign functions is seamless on the \CDSL side, but
naturally has requirements on the corresponding C code. The C side must
respect the \CDSL type system, and, for example, keep all shared state
internal to the abstract type to comply with linearity constraints. It must
also be terminating and implement the user-supplied semantics that appear in
the corresponding shallow embedding of the \CDSL program in Isabelle/HOL ---
ideally the user should provide a formal proof to discharge the corresponding
assumption of the compiler certificate theorem.

Abstract functions can be higher-order and provide the
iteration constructs that are intentionally left out from core \cdsl.
E.g.\ line 21, \cdslfunc{uarray\_map\_no\_break} implements a map iterator
for arrays. In our
file system applications we have found it sufficient to provide a small
library of iterators for types such as arrays. We also interfaced to an
existing mature red-black tree implementation.

Returning to the example in \autoref{fig:cdsl-snippet}, lines 3--7 show basic
type constructors and declarations of 
variants, records and tuples
using type variables and the primitive type \cdsltype{U32}. For instance, type
\cdsltype{Cnt} is defined as a pair of \cdsltype{UArray Node} and a function
type. Types in \cdsl are structural~\citep{Pierce_02}, i.e.\ two types with
the same structure but different names are intensionally equal.

Moreover, line 17 calls the abstract polymorphic function   
\cdslfunc{uarray\_create}, instantiated with type argument \cdsltype{Node}.
The \code{!nd} notation temporarily turns a linear object of type
\cdsltype{Node} into a read-only one (see \autoref{s:letbang}). The two
basic, non-linear fields \code{to} and \code{fr} in type \cdsltype{Node} can
directly be accessed read-only using projection functions.
Line 18 and 29 are pattern matches on the result of the function invocation.
Line 19 shows surface syntax for \CDSL's linear \textbf{take} construct (see
\autoref{sec:records}), accessing and binding the \code{mbuf} field of
\code{nd} to the name~\code{mbuf} (punning as in Haskell), as well as
binding the rest of the record to the name \code{nd\_t}.

The linear type system tracks that the field \code{mbuf} is logically absent
in \code{nd\_t}. It also tracks that \code{nd} on line 19 has been used,
so cannot be accessed again. Thus the programmer is safe to bind a new object
to the same name \code{nd} (on line 26) without worrying about name
shadowing. Line 26 shows surface syntax for $\PUT$, the dual to $\TAKE$,
which re-establishes the \code{mbuf} fields in the example.

\subsection{Types and Kinding}\label{s:kinding}

\citet{Wadler_90} first noted that linear types can be used as a way to safely
model mutable state and similar effects while maintaining a purely functional
semantics.  \citet{Hofmann_00} later proved Wadler's intuition by
showing that, for a linear language, imperative C code can implement a simple
set-theoretic semantics. We use linear types for two reasons: to ensure safe handling of heap-allocated objects, without the need for runtime 
support, and to allow us to assign to \CDSL programs a simple, equational, purely functional semantics implemented via mutable state and imperative effects.

\begin{figure}
\begin{grammar}
\text{prim. types}     & t              & \Coloneqq & \PrimType{U8} \alt \PrimType{U16} \alt \PrimType{U32} \alt \PrimType{U64} \alt \PrimType{Bool} \\
\text{types}           & \tau, \rho     & \Coloneqq & \alpha \alt \Observed{\alpha} \alt \Unit \\
                       &                & \alt      & t \alt \AbsTy{T}{\many{\tau}}{m} \alt \FunTy{\tau}{\rho} \\
                       &                & \alt      & \VariantTy{\many{\Cons{C}{\tau}}} \alt \RecordTy{\many{\FieldTy{f}{\perhaps{\tau}}}}{m} \\
\text{field types}     & \perhaps{\tau} & \Coloneqq & \tau \alt \taken{\tau}\\
\text{permissions}     & \mathcal{P}    & =         & \{ \Discardable, \Shareable, \Escapable \}\\
\text{kinds}           & \kappa         & \subseteq & \mathcal{P} \\
\text{polytypes}       & \pi            & ::=       & \PolyTy{\many{\OfKind{\alpha}{\kappa}}}{\tau}\\
\text{modes}           & m              & \Coloneqq & \ReadOnly \alt \Writable \alt \Unboxed \\
\text{type variables}  &                & \ni       & \alpha, \beta \\
\text{abs. type names} &                & \ni       & \AbsN{T}, \AbsN{U} \\
\text{kind context}    & \Delta         & \Coloneqq & \many{\ofKind{\alpha}{\kappa}} \\
\text{type context}    & \Gamma         & \Coloneqq & \many{\ofType{x}{\tau}}
\end{grammar} \\[-1em]
\begin{tabular}{p{0.24\columnwidth}p{0.7\columnwidth}}
\boxlabel{$\Weakening{\Delta}{\Gamma_1}{\Gamma_2}$} &
\begin{displaymath}
  \inferrule{\text{for each $i$:}\ \Kinding{\Delta}{\tau_i}{\{\Discardable\}}}
            {\Weakening{\Delta}{\many{\ofType{x_i}{\tau_i}},\Gamma}{\Gamma}}
\end{displaymath} \\[-1.3em]
\boxlabel{$\Contraction{\Delta}{\Gamma_1}{\Gamma_2}{\Gamma_3}$} &
\begin{displaymath}
  \inferrule{\text{for each $i$:}\ \Kinding{\Delta}{\tau_i}{\{\Shareable\}}}
            {\Contraction{\Delta}{\many{\ofType{x_i}{\tau_i}},\Gamma_1,\Gamma_2}{\many{\ofType{x_i}{\tau_i}},\Gamma_1}{\many{\ofType{x_i}{\tau_i}},\Gamma_2}}
\end{displaymath}\end{tabular}
\vspace{-0.7em}
\begin{center}
  ($\many{\text{overbar}}$ indicates lists, i.e. zero or more)
\end{center}
\vspace{-0.5em}
\caption{Type Structure of \CDSL \& structural context operations}
\label{fig:typestruct}
\end{figure}

The type structure and associated syntax of \CDSL is presented in \autoref{fig:typestruct}. Our type system is loosely based on the polymorphic
$\lambda_{\text{URAL}}$ of \citet{Ahmed_FM_05}. We restrict this polymorphism to be rank-1 and predicative, in the style of ML, to permit easy implementation
by specialisation with minimal performance penalty. 

To ease implementation, and to eliminate any direct dependency on a heap allocator, we require that all functions be defined on the top-level. 
This eliminates the need for linear function types: any top-level function can be shared freely because they cannot capture \emph{any} local variables, 
let alone linear ones.

We include a set of primitive integer types ($\PrimType{U8}$, $\PrimType{U16}$
etc.). Records $\RecordTy{\many{\FieldTy{f}{\perhaps{\tau}}}}{m}$ comprise
(1)~a sequence of fields $f :: \perhaps{\tau}$, where $\taken{\tau}$ is the type on
an inaccessible field, and (2)~a mode $m$ (see
 \mbox{\autoref{sec:records}} and \autoref{s:kindrec} for a more detailed description). We also have polymorphic variants
$\VariantTy{\many{\Cons{C}{\tau}}}$, a generalised sum type in the style of
OCaml, the mechanics of which are briefly described in
\autoref{s:variants}. Abstract types $\AbsTy{T}{\many{\tau}}{m}$ are also
parametrised by modes.
We omit product types from this presentation; 
they are desugared into unboxed records.

The most obvious similarity to $\lambda_{\text{URAL}}$ is our use of \emph{kinds} to determine if a type may be freely shared or discarded, as 
opposed to earlier linear type systems, such as that of~\citet{Wadler_90}, where a type's linearity is encoded directly into its syntactic structure. Kinds 
in \CDSL are sets of \emph{permissions}, denoting whether a variable of that type may be discarded without being used ($\Discardable$), shared freely and used 
multiple times ($\Shareable$), or safely bound in a $\LET!$ expression ($\Escapable$). A \emph{linear} type, values of which must be used exactly once, 
has a kind
that excludes~$\Discardable$ and~$\Shareable$, and so forbids it being discarded or shared.
We discuss $\LET!$ expressions in \autoref{sec:kindletb}. 

Another similarity to $\lambda_{\text{URAL}}$ is that we explicitly represent the context operations of weakening and contraction, normally relegated to structural rules, 
as explicit judgements: $\Weakening{\Delta}{\Gamma}{\Gamma'}$ for weakening (discarding assumptions) and $\Contraction{\Delta}{\Gamma}{\Gamma_1}{\Gamma_2}$ for contraction (duplicating them). 
The rules for these judgements are presented in \autoref{fig:typestruct}. For a typing assumption to be discarded (respectively duplicated),
the type must have kind $\{\Discardable\}$ (resp. $\{\Shareable\}$).

\begin{figure}
\begin{inductive}{\Kinding{\Delta}{\tau}{\kappa}}
  \inferrule{ }{\Kinding{\Delta}{\Unit}{\kappa}}{\rulename{KUnit}} \quad
  \inferrule{ }{\Kinding{\Delta}{t}{\kappa}}{\rulename{KPrim}} \quad
  \inferrule{ }{\Kinding{\Delta}{\FunTy{\tau}{\rho}}{\kappa}}{\rulename{KFun}} \\
  \inferrule{(\ofKind{\alpha}{\kappa'}) \in \Delta \quad \kappa \subseteq \kappa'}
            {\Kinding{\Delta}{\alpha}{\kappa}}{\rulename{KVar}} \quad
  \inferrule{(\ofKind{\alpha}{\kappa'}) \in \Delta \quad \kappa \subseteq \BangF{\kappa'}}
            {\Kinding{\Delta}{\alpha!}{\kappa}}{\rulename{KVar}!} \\
  \inferrule{ \text{for each $i$:}\ \Kinding{\Delta}{\tau_i}{\kappa} }
            {\Kinding{\Delta}{\VariantTy{\many{\Cons{C_\mathit{i}}{\tau_i}}}}{\kappa}}{\rulename{KVariant}} \\
  \inferrule{ \ofKind{m}{\kappa'} \\ \kappa \subseteq \kappa' \\\\ \text{for each $i$:}\ \Kinding{\Delta}{\tau_i}{\kappa} }
            {\Kinding{\Delta}{\AbsTy{T}{\many{\tau_i}}{m}}{\kappa}}{\rulename{KAbs}} \quad\!\!\!
  \inferrule{ \ofKind{m}{\kappa'} \\ \kappa \subseteq \kappa' \\\\ \text{for each $\tau_i$ not taken:}\ \Kinding{\Delta}{\tau_i}{\kappa} }
            {\Kinding{\Delta}{\RecordTy{\many{\FieldTy{f$_i$}{\perhaps{\tau_i}}}}{m}}{\kappa}}{\rulename{KRec}} \\
\end{inductive}
\vspace{-2.5em}
\begin{inductive}{\ofKind{m}{\kappa}}
  \inferrule{}{\ofKind{\ReadOnly}{\{\Discardable, \Shareable\}}} \quad
  \inferrule{}{\ofKind{\Writable}{\{\Escapable\}}} \quad
  \inferrule{}{\ofKind{\Unboxed}{\{\Discardable, \Shareable, \Escapable\}}} 
\end{inductive}
\vspace{-1em}
  \boxlabel{$\BangF{\cdot} : \tau \rightarrow \tau$}
  \begin{displaymath}
    \begin{array}{lcl}
      \BangF{\alpha}                                       & = & \Observed{\alpha} \\
      \BangF{\Observed{\alpha}}                            & = & \Observed{\alpha} \\
      \BangF{\Unit}                                        & = & \Unit \\
      \BangF{t}                                            & = & t \\
      \BangF{\AbsTy{T}{\many{\tau_i}}{m}}                  & = & \AbsTy{T}{\many{\BangF{\tau_i}}}{\BangF{m}} \\
      \BangF{\FunTy{\tau}{\rho}}                           & = & \FunTy{\tau}{\rho} \\
      \BangF{\VariantTy{\many{\Cons{C_i}{\tau_i}}}}        & = & \VariantTy{\many{\Cons{C_i}{\BangF{\tau_i}}}} \\
      \BangF{\RecordTy{\many{\FieldTy{f$_i$}{\perhaps{\tau_i}}}}{m}} & = & \RecordTy{\many{\FieldTy{f$_i$}{\BangF{\perhaps{\tau_i}}}}}{\BangF{m}} \\
    \end{array}
  \end{displaymath}
  \boxlabel{$\BangF{\cdot} : \kappa \rightarrow \kappa$}
  \begin{displaymath}
     \BangF{\kappa} \quad = \quad \begin{cases}
                                     \kappa                        & \text{if}\ \{\Discardable, \Shareable\} \subseteq \kappa \\
                                     \{\Discardable, \Shareable\}  & \text{otherwise}
                                  \end{cases}
  \end{displaymath}
  \boxlabel{$\BangF{\cdot} : m \rightarrow m$}
  \begin{displaymath}
    \begin{array}{lcl}
      \BangF{\ReadOnly}                                       & = & \ReadOnly\\
      \BangF{\Writable}                                       & = & \ReadOnly\\
      \BangF{\Unboxed}                                        & = & \Unboxed \\
    \end{array}
  \end{displaymath}
\caption{Kinding rules for \CDSL types and the $\BangF{\cdot}$ operator}
\label{fig:kinding}
\end{figure}

The full kinding rules for the types of \CDSL are given in \autoref{fig:kinding}. Basic types such as $\Unit$ or $\PrimType{U8}$, as well as functions,
are simply passed by value and do not contain any heap references, so they may be given any kind. Kinding for structures and abstract functions is discussed
shortly in \autoref{s:kindrec}.

A type may have multiple kinds, as a nonlinear type assumption may be used linearly, never being shared and being used exactly once. Therefore, a type with a
permissive kind, such as $\{\Discardable, \Shareable\}$, would be an acceptable instantiation of a type variable of kind $\emptyset$, as we are free to 
\emph{waive} permissions that are included in a kind. We can prove formally by straightforward rule induction:
\begin{lemma}[Waiving rights] If $\Kinding{\Delta}{\tau}{\kappa}$ and $\kappa' \subseteq \kappa$, then $\Kinding{\Delta}{\tau}{\kappa'}$.
\end{lemma}
\noindent This result allows for a simple kind-checking algorithm, not immediately apparent from the rules. For example, the maximal kind of an unboxed
structure with two fields of type $\tau_1$ and $\tau_2$ respectively can be computed by taking the intersection of the computed maximal kinds of $\tau_1$ and $\tau_2$. This result ensures
that this intersection is also a valid kind for $\tau_1$ and $\tau_2$.

\subsubsection{Kinding for Records and Abstract Types}\label{s:kindrec}

Recall that \CDSL may be extended with \emph{abstract types}, implemented in
C, which we write as
$\AbsTy{T}{\many{\tau_i}}{m}$ in our formalisation. We allow abstract types
to take any number of \emph{type parameters} ${\tau_i}$, where each specific
instance corresponds to a distinct C type. For example, a $\mathtt{List}$ abstract type,
parameterised by its element type, would correspond to a family of C $\mathtt{List}$ types,
each one specialised to a particular concrete element type. Because the implementations of these
types are user supplied, the user is free to specialise implementations based on these type parameters,
for example representing an array of boolean values as a bitstring, so long as they can show 
that every different operation implementation is a refinement of the same user-supplied CDSL 
semantics for that operation.

Values of abstract types may be represented by references to heap data structures. 
Specifically, an abstract type or structure is stored on the heap when its associated
\emph{storage mode}~$m$ is not ``\Unboxed''. For boxed records and abstract
types, the storage mode distinguishes between those that are ``\Writable'' vs.
``\ReadOnly''. The same is true for record types, written $\RecordTy{\many{\FieldTy{f}{\perhaps{\tau}}}}{m}$,
which are discussed in more detail in \autoref{sec:records}.

The storage mode $m$ affects the maximal kind that can be assigned
to the type. For example, an unboxed structure with two components of type $\PrimType{U8}$ is freely shareable, but if the structure is
instead stored on the heap, then a writable reference to that structure must be linear. Thus, the type given to such references has the ``$\Writable$'' mode,
whose kind is $\{\Escapable\}$, thereby preventing such a reference from being assigned a nonlinear kind such as $\{\Discardable,\Shareable\}$.


\subsubsection{Kinding and $\textbf{bang}$}
\label{sec:kindletb}

Like \citet{Wadler_90}, we allow linear values to be shared read-only in a limited scope. This is useful for practical programming
in a language with linear types, as it makes our types more informative. For example, to write a function to determine the size of a (linear) buffer object, 
a naive approach would be to write a function:
$$\mathsf{size} : \mathtt{Buf} \rightarrow \mathtt{U32} \times \mathtt{Buf}$$
This function has a cumbersome additional return value just so that the linear
argument is not discarded. Further, the type above does not express the fact that the
input buffer and output buffer are identical --- this would need to be established by additional proof.  To address this problem, we include a type operator~$\BangF{\cdot}$, in the style
of Wadler's $!$ operator, which changes all writable modes in a type to read-only ones. The full definition of $\BangF{\cdot}$ is in \autoref{fig:kinding}.
We can therefore write the type of our function as:
$$\mathsf{size} : \BangF{\mathtt{Buf}} \rightarrow \mathtt{U32}$$
For any valid type $\tau$, the kind of $\BangF{\tau}$ will be
nonlinear, which means that our $\mathsf{size}$ function no longer needs to be encumbered by the 
extra return value. This kinding result is formally stated as:

\begin{lemma}[Kinding for $\BangF{\cdot}$] For any type $\tau$, if $\Kinding{\Delta}{\tau}{\kappa}$ then $\Kinding{\Delta}{\BangF{\tau}}{\BangF{\kappa}}$.
\end{lemma}

\noindent To integrate this type operator with parametric polymorphism, we borrow a trick from Odersky's Observer types~\citep{Odersky_92}, and tag type variables that
have been made read only, using the syntax $\alpha!$. Whenever a variable $\alpha$ is instantiated to some concrete type $\tau$, we also replace 
$\alpha!$ with $\BangF{\tau}$. The lemma above ensures that our kinding rule for such tagged variables is sound, and enables us to prove the following:

\begin{lemma}[Type instantiation preserves kinds] For any type $\tau$, $\Kinding{\many{\ofKind{\alpha_i}{\kappa_i}}}{\tau}{\kappa}$
      implies $\Kinding{\Delta}{\Subst{\tau}{\many{\alpha_i}}{\many{\rho_i}}}{\kappa}$ when, for each $i$, $\Kinding{\Delta}{\rho_i}{\kappa_i}$.
\end{lemma}

\subsection{Expressions and Typing}
\begin{figure}
\begin{grammar}
\text{primops}         & o              & \in       & \{\texttt{+}, \texttt{*}, \texttt{/}, \texttt{<=}, \texttt{==}, \texttt{||}, \texttt{{<}<}, \dots\} \\
\text{literals}        & \ell           & \in       & \{123, \mathtt{True}, \texttt{'a'}, \dots\} \\
\text{expressions}     & e              & \Coloneqq & \VarN{x} \alt \Unit \alt \TyApp{f}{\many{\tau}} \alt \GenPrimOp{o}{\many{e}} \alt \App{e_1}{e_2} \\
                       &                & \alt      & \Let{x}{e_1}{e_2} \\
                       &                & \alt      & \LetBang{\many{y}}{x}{e_1}{e_2} \\
                       &                & \alt      & \If{e_1}{e_2}{e_3} \\
                       &                & \alt      & \ell \alt \Cast{t}{e} \alt \Promote{\many{\Cons{C}{\tau}}}{e} \\
                       &                & \alt      & \Case{e_1}{\Cons{C}{\VarN{x}}}{e_2}{y}{e_3} \\
                       &                & \alt      & \Esac{e} \alt \Cons{C}{e} \\
                       &                & \alt      & \StructInit{\many{\FieldEq{f}{e}}} \alt \Member{e}{f} \alt \Put{e_1}{f}{e_2} \\
                       &                & \alt      & \Take{x}{f}{y}{e_1}{e_2} \\
\text{function def.}   & d              & \Coloneqq & \FunDef{f}{\pi}{x}{e} \alt \AbsFunDef{f}{\pi} \\
\text{programs}        & P              & \Coloneqq & \many{d}\\
\text{function names}  &                & \ni       & \FunN{f}, \FunN{g} \\
\text{variables}       &                & \ni       & \VarN{x}, \VarN{y} \\
\text{constructors}    &                & \ni       & \ConsN{A},\ConsN{B},\ConsN{C} \\
\text{record fields}   &                & \ni       & \FieldN{f}, \FieldN{g} \\
\end{grammar}
\begin{displaymath}
  \begin{array}{lclr}
    \PrimOpType{\cdot} & : & o \rightarrow \many{t} \times t & \text{(primop types)} \\
    \FunDefn{\cdot} &  : & f \rightarrow d  & \text{(definition environment)}\\
    |\cdot| & : & t \rightarrow \mathbb{N} & \text{(maximum value)}  \\
  \end{array}
\end{displaymath}
\caption{Syntax of \CDSL programs (after desugaring)}
\label{fig:syntax}
\end{figure}
\noindent 
While \CDSL features a rich surface syntax, due to space constraints, we only document the (full) core language in \autoref{fig:syntax} to which the surface syntax is desugared.

\begin{figure*}
  \begin{inductive}{\Typing{\Delta}{\Gamma}{e}{\tau}}
    \inferrule{\Weakening{\Delta}{\Gamma}{\ofType{x}{\tau}}}{\Typing{\Delta}{\Gamma}{x}{\tau}}{\rulename{Var}} \quad
    \inferrule{ }{\Typing{\Delta}{\Gamma}{\Unit}{\Unit}}{\rulename{Unit}} \quad
    \inferrule{ \ell < |t|}{\Typing{\Delta}{\Gamma}{\ell}{t}}{\rulename{Literal}} \quad
    \inferrule{ \TypingS{\Delta}{\Gamma}{\many{e_i}}{\many{t_i}} \\\\
               \PrimOpType{o} = (\many{t_i}, t) }
              {\Typing{\Delta}{\Gamma}{\GenPrimOp{o}{\many{e_i}}}{t}}{\rulename{PrimOp}}\quad
    \inferrule{ \Typing{\Delta}{\Gamma}{e}{t'} \\ |t'| \le |t| }{\Typing{\Delta}{\Gamma}{\Cast{t}{e}}{t}}{\rulename{Cast}} \\
    \inferrule{\Contraction{\Delta}{\Gamma}{\Gamma_1}{\Gamma_2} \\\\
               \Typing{\Delta}{\Gamma_1}{e_1}{\rho \rightarrow \tau} \\
               \Typing{\Delta}{\Gamma_2}{e_2}{\rho} }
              {\Typing{\Delta}{\Gamma}{\App{e_1}{e_2}}{\tau}}{\rulename{App}} \quad\!\!
    \inferrule{\FunTyEnv{f}{\PolyTy{\many{\OfKind{\alpha_i}{\kappa_i}}}{\tau \rightarrow \tau'}}\\\\
               \text{for each $i$:}\ \Kinding{\Delta}{\rho_i}{\kappa_i}}
              {\Typing{\Delta}{\Gamma}{\TyApp{f}{\many{\rho_i}}}{\Subst{(\tau \rightarrow \tau')}{\many{\alpha_i}}{\many{\rho_i}}}}{\rulename{Fun}} \quad\!\!
    \inferrule{\Contraction{\Delta}{\Gamma}{\Gamma_1}{\Gamma_2} \\
               \Typing{\Delta}{\Gamma_1}{e_1}{\VariantTy{\Cons{A}{\rho} \alt \many{\Cons{C_i}{\tau_i}}}} \\\\
               \Typing{\Delta}{\ofType{x}{\rho}, \Gamma_2}{e_2}{\tau} \\
               \Typing{\Delta}{\ofType{y}{\VariantTy{\many{\Cons{C_i}{\tau_i}}}}, \Gamma_2}{e_3}{\tau}}
              {\Typing{\Delta}{\Gamma}{\Case{e_1}{\Cons{A}{\VarN{x}}}{e_2}{y}{e_3}}{\tau}}{\rulename{Case}}    
  \\
  \end{inductive}
  \begin{center}
    \begin{tabular}{p{0.625\textwidth}|p{0.25\textwidth}}
      \begin{inductive0}
    \begin{tabular}{c}
    \inferrule{\Typing{\Delta}{\Gamma}{e}{\tau}}
              {\Typing{\Delta}{\Gamma}{\Cons{C}{e}}{\VariantTy{\Cons{C}{\tau}}}}{\rulename{Cons}} \quad\!\!\!
    \inferrule{\Typing{\Delta}{\Gamma}{e}{\VariantTy{\many{\Cons{B}{\rho}}}} \\
               \many{\Cons{B}{\rho}} \subseteq \many{\Cons{C}{\tau}}}
              {\Typing{\Delta}{\Gamma}{\Promote{\many{\Cons{C}{\tau}}}{e}}{\VariantTy{\many{\Cons{C}{\tau}}}}}{\rulename{Prom}} \quad\!\!\!
    \inferrule{\Typing{\Delta}{\Gamma}{e}{\VariantTy{\Cons{C}{\tau}}}}
              {\Typing{\Delta}{\Gamma}{\Esac{e}}{\tau}}
              {\rulename{Esac}} \quad\!\!\!
\\\\
        \inferrule{\Contraction{\Delta}{\Gamma}{\Gamma_1}{\Gamma_2} \\\\
                   \Typing{\Delta}{\Gamma_1}{e_1}{\rho} \\
                   \Typing{\Delta}{\ofType{x}{\rho}, \Gamma_2}{e_2}{\tau}}
                  {\Typing{\Delta}{\Gamma}{\Let{x}{e_1}{e_2}}{\tau}}{\rulename{Let}} \quad
        \inferrule{\Contraction{\Delta}{\Gamma}{\Gamma_1}{\Gamma_2} \\
                   \Kinding{\Delta}{\rho}{\{\Escapable\}}\\\\
                   \Typing{\Delta}{\many{\ofType{v_i}{\BangF{\tau_i}}}, \Gamma_1}{e_1}{\rho} \\\\
                   \Typing{\Delta}{\many{\ofType{v_i}{\tau_i}}, \ofType{x}{\rho},\Gamma_2}{e_2}{\tau}}
                  {\Typing{\Delta}{\many{\ofType{v_i}{\tau_i}}, \Gamma}{\LetBang{\many{v_i}}{x}{e_1}{e_2}}{\tau}}{\rulename{Let}!} \\
        \end{tabular}
      \end{inductive0}
      &\vspace{-2ex}
      \begin{inductive}{\TypingS{\Delta}{\Gamma}{\many{e}}{\many{\tau}}}
        \inferrule{\Weakening{\Delta}{\Gamma}{\emptyset} }
                  {\TypingS{\Delta}{\Gamma}{\varepsilon}{\varepsilon}}{\rulename{L}_\varepsilon} \\
        \inferrule{\Contraction{\Delta}{\Gamma}{\Gamma_1}{\Gamma_2} \\\\
                   \Typing{\Delta}{\Gamma_1}{e}{\tau} \\
                   \TypingS{\Delta}{\Gamma_2}{\many{e_i}}{\many{\tau_i}}}
                  {\TypingS{\Delta}{\Gamma}{e\ \many{e_i}}{ \tau\ \many{\tau_i} }}{\rulename{L}_C}
      \end{inductive}
    \end{tabular}
  \end{center}\vspace{1.5ex}
  \begin{inductive0}
    \inferrule{\Contraction{\Delta}{\Gamma}{\Gamma_1}{\Gamma_2} \\
               m \neq \ReadOnly \\\\
               \Typing{\Delta}{\Gamma_1}{e_1}{\RecordTy{\many{\FieldTy{g$_i$}{\perhaps{\tau_i}}}, \FieldTy{f}{\rho} ,\many{\FieldTy{g$_j$}{\perhaps{\tau_j}}}}{m}} \\\\
               \Typing{\Delta}{\ofType{x}{\RecordTy{\many{\FieldTy{g$_i$}{\perhaps{\tau_i}}}, \FieldTy{f}{\taken{\rho}} ,\many{\FieldTy{g$_j$}{\perhaps{\tau_j}}}}{m}}, \ofType{y}{\rho}, \Gamma_2}{e_2}{\tau}}
              {\Typing{\Delta}{\Gamma}{\Take{x}{f}{y}{e_1}{e_2}}{\tau}}{\rulename{Take}_1} \quad
    \inferrule{\Contraction{\Delta}{\Gamma}{\Gamma_1}{\Gamma_2} \\
               \Kinding{\Delta}{\rho}{\{\Shareable\}} \\\\
               m \neq \ReadOnly \\ \perhaps{\tau_k} = \rho \\
               \Typing{\Delta}{\Gamma_1}{e_1}{\RecordTy{\many{\FieldTy{f$_i$}{\perhaps{\tau_i}}}}{m}} \\\\
               \Typing{\Delta}{\ofType{x}{\RecordTy{\many{\FieldTy{f$_i$}{\perhaps{\tau_i}}}}{m}}, \ofType{y}{\rho}, \Gamma_2}{e_2}{\tau}}
              {\Typing{\Delta}{\Gamma}{\Take{x}{f$_k$}{y}{e_1}{e_2}}{\tau}}
              {\rulename{Take}_2} \\
    \inferrule{\Contraction{\Delta}{\Gamma}{\Gamma_1}{\Gamma_2} \\
               m \neq \ReadOnly \\\\
               \Typing{\Delta}{\Gamma_1}{e_1}{\RecordTy{\many{\FieldTy{g$_i$}{\perhaps{\tau}_i}}, \FieldTy{f}{\taken{\rho}} ,\many{\FieldTy{g$_j$}{\perhaps{\tau}_j}}}{m}} \\
               \Typing{\Delta}{\Gamma_2}{e_2}{\rho}}
              {\Typing{\Delta}{\Gamma}{\Put{e_1}{f}{e_2}}{\RecordTy{\many{\FieldTy{g$_i$}{\perhaps{\tau}_i}}, \FieldTy{f}{\rho} ,\many{\FieldTy{g$_j$}{\perhaps{\tau}_j}}}{m}}}{\rulename{Put}_1} \quad
    \inferrule{\Contraction{\Delta}{\Gamma}{\Gamma_1}{\Gamma_2} \\
               m \neq \ReadOnly \\ \perhaps{\tau}_k = \rho\\\\
               \Typing{\Delta}{\Gamma_1}{e_1}{\RecordTy{\many{\FieldTy{f$_i$}{\perhaps{\tau}_i}}}{m}} \\
               \Kinding{\Delta}{\rho}{\{\Discardable\}} \\
               \Typing{\Delta}{\Gamma_2}{e_2}{\rho}}
              {\Typing{\Delta}{\Gamma}{\Put{e_1}{f$_k$}{e_2}}{\RecordTy{\many{\FieldTy{f$_i$}{\perhaps{\tau}_i}}}{m}}}{\rulename{Put}_2} \\
    \inferrule{\Kinding{\Delta}{\RecordTy{\many{\FieldTy{g$_i$}{\perhaps{\rho}_i}}, \FieldTy{f}{\tau} ,\many{\FieldTy{g$_j$}{\perhaps{\rho}_j}}}{m}}{\{\Shareable\}} \\\\
              \Typing{\Delta}{\Gamma_1}{e_1}{\RecordTy{\many{\FieldTy{g$_i$}{\perhaps{\rho}_i}}, \FieldTy{f}{\tau} ,\many{\FieldTy{g$_j$}{\perhaps{\rho}_j}}}{m}}}
              {\Typing{\Delta}{\Gamma}{\Member{e}{f}}{\tau}}{\rulename{Member}} \quad
    \inferrule{\TypingS{\Delta}{\Gamma}{\many{e_i}}{\many{\tau_i}}}
              {\Typing{\Delta}{\Gamma}{\StructInit{\many{\FieldEq{f$_i$}{e_i}}}}{\RecordTy{\many{\FieldTy{f$_i$}{\tau_i}}}{\Unboxed}}}
              {\rulename{Struct}}
  \end{inductive0}
  \caption{Typing rules for \CDSL}
  \label{fig:typing}
\end{figure*}

\autoref{fig:typing} shows the typing rules for \CDSL expressions.
Many of these are standard for any linear type system. We will
discuss here the rules for $\LET!$, where we have taken a slightly different approach to established literature, and the rules for the 
extensions we have made to the type system, such as variants and record types.

\subsubsection{Typing for $\LET!$}\label{s:letbang}

On the expression level, the programmer can use $\LET!$ expressions, in the style of \citet{Wadler_90}, to temporarily convert variables of linear types to their
read-only equivalents, allowing them to be freely shared. In this example, we wish to copy a buffer $b_2$ onto a buffer $b_1$ only when $b_2$ will fit inside
$b_1$. 
$$\begin{array}{l}
  \LetBang{ b_1, b_2 }{\mathit{ok}}{(\mathsf{size}(b_2) < \mathsf{size}(b_1))}{ } \\
  \quad\If{\mathit{ok}}{\mathsf{copy}(b_1, b_2)}{\ldots}
  \end{array}$$ 
Note that even though $b_1$ and $b_2$ are used multiple times, they are only used once in a linear context. Inside the $\LET!$ binding, they have been made
temporarily nonlinear. Our kind system ensures these read-only, shareable references inside $\LET!$ bindings cannot ``escape'' into
the outside context. For example, the expression $\LetBang{b}{b'}{b}{\mathsf{copy}(b, b')}$ would violate the invariants of the linear type system, and ruin the purely
functional abstraction that linear types allow, as both $b$ and $b'$ would refer to the same object, and a destructive update to $b$ would change the shareable $b'$. 

We are able to use the existing kind system to handle these safety checks with
the inclusion of the $\Escapable$ permission, for
$\Escapable$scapable, which indicates that the type may be safely returned
from within a $\LET!$. We ensure, via the typing rules of
\autoref{fig:typing}, that the left hand side of the binding ($\mathit{ok}$ in the example) has the
$\Escapable$ permission, which excludes temporarily nonlinear references via
$\BangF{\cdot}$ (see \autoref{fig:kinding}). Our solution is as powerful as
Odersky's, but we encode the restrictions in the kind system directly, not as 
side-condition constraints that recursively descend into the structure of
the binding's type.

%
%

\subsubsection{Typing for Variants}\label{s:variants}
A variant type $\VariantTy{\many{\Cons{C_i}{\tau_i}}}$ is a generalised sum type, where each alternative is distinguished by a unique \emph{data constructor}~$\ConsN{C_i}$. The order in which the constructors appear in the type is not important.
One can create a variant type with a single alternative simply by invoking a constructor, e.g. $\Cons{Some}{255}$ might be given the type
$\VariantTy{\Cons{Some}{\PrimType{U8}}}$. The original value of $255$ can be retrieved using the $\ESAC$ construct. The set of alternatives is enlarged 
by using $\PROMOTE$ expressions that are automatically inserted by the type-checker of the surface language, which uses subtyping to infer the type of a given
variant. A similar trick is used for numeric literals and $\CAST$. 

In order to pattern match on a variant, we provide a $\CASE$ construct that attempts to match against one constructor. If the constructor does not match, 
it is \emph{removed} from the type and the reduced type is provided to the $\ELSE$ branch. In this way, a traditional multi-way pattern match can be desugared
by nesting:
\begin{displaymath}
\begin{array}{@{}lcl@{}}
 \begin{array}{l}
   \CASE\ x\ \OF \\
  \quad \Cons{A}{a} \rightarrow e_a \\
  \quad \Cons{B}{b} \rightarrow e_b \\
  \quad \Cons{C}{c} \rightarrow e_c 
 \end{array}  & \text{becomes} & \begin{array}{l}
   \CASE\ x\ \OF \\
   \quad \Cons{A}{a} \rightarrow {e_a} \\
   \quad \ELSE\ x' \rightarrow \CASE\ x'\ \OF \\
   \quad\quad \Cons{B}{b} \rightarrow e_b \\
   \quad\quad \ELSE\ x'' \rightarrow \LET\ c = \Esac{x''}\ \IN\ e_c
 \end{array} 
\end{array}
\end{displaymath}
Note that because the typing rule for $\ESAC$ only applies when only one alternative remains, our pattern matching is necessarily total.

\subsubsection{Typing for Records}
\label{sec:records}


Some care is needed to reconcile record types and linear types.
Assume that $\mathtt{Object}$ is a type synonym for an (unboxed) record type containing an integer and two (linear) buffers.
$$ \mathtt{Object} = \{ \text{size} :: \mathtt{U32}, \text{b}_1 :: \mathtt{Buf}, \text{b}_2 :: \mathtt{Buf} \}\ \Unboxed$$
Let us say we want to extract the field $\text{b}_1$ from an $\mathtt{Object}$. If we extract just a single $\mathtt{Buf}$, we have implicitly discarded the other buffer $\text{b}_2$.
But, we can't return the entire $\mathtt{Object}$ along with the $\mathtt{Buf}$, as this would introduce aliasing. Our solution is to return along with the $\mathtt{Buf}$
 an $\mathtt{Object}$ where the field $\text{b}_1$ cannot be extracted again, and reflect this in the field's type, written as $\text{b}_1 :: \taken{\mathtt{Buf}}$. 
This field extractor, whose general form is $\Take{x}{f}{y}{e_1}{e_2}$, operates as follows: given a record~$e_1$, it binds the field $f$ of $e_1$ 
to the variable
$y$, and the new record to the variable $x$ in $e_2$. Unless the type of the field $f$ has kind $\{\Shareable\}$, that field will be marked as unavailable, or \emph{taken}, in the type of the new record $x$.

Conversely, we also introduce a $\PUT$ operation, which, given a record with a taken field, allows a new value to be supplied in its place. The expression $\Put{e_1}{f}{e_2}$ returns the
record in $e_1$ where the field $f$ has been replaced with the result of $e_2$. Unless the type of the field $f$ has kind $\{\Discardable\}$, that field must already be taken, to avoid
accidentally destroying our only reference to a linear resource.

Unboxed records can be created using a simple struct literal $\StructInit{\many{\FieldEq{f$_i$}{e_i}}}$. We also allow records to be stored on the heap to minimise unnecessary copying,
as unboxed records are passed by value.  These boxed records are created by invoking an externally-defined C allocator function.
For these allocation functions, it is often convenient to allocate a record with all fields already taken, to indicate that they are uninitialised. Thus a function for allocating
\texttt{Object}-like records might return values of type: $ \{ \text{size} :: \taken{\mathtt{U32}}, \text{b}_1 :: \taken{\mathtt{Buf}}, \text{b}_2 :: \taken{\mathtt{Buf}} \}\ \Writable$.

For any nonlinear record (that is, (1)~read-only boxed records, which cannot have linear fields, as well as~(2)~unboxed records without linear fields)
 we also allow traditional member syntax $\Member{e}{f}$ for field access. The typing rules for all of these expressions are given in \autoref{fig:typing}. 

\subsubsection{Type Specialisation} 
\label{sec:typing:poly}
As mentioned earlier, we implement parametric polymorphism by specialising code to avoid paying the performance penalties of other approaches such as
boxing. This means that polymorphism in our language is restricted to predicative rank-1 quantifiers.

This allows us to specify dynamic objects, such as our value typing relations (see \autoref{sec:updvalrel}) and our dynamic semantics (see \autoref{sec:dynsem}), in
terms of simple monomorphic types, without type variables. Thus, in order to evaluate a polymorphic program, each type variable must first be instantiated
to a monomorphic type. We show that typing of the instantiated program follows from the typing of the polymorphic program, if the type instantiation used matches
the kinds of the type variables.

\begin{lemma}[Type specialisation]
\label{lemma:spec} 
$\Typing{\many{\ofKind{\alpha_i}{\kappa_i}}}{\Gamma}{e}{\tau}$
      implies $\Typing{\Delta}
                      {\Subst{\Gamma}{\many{\alpha_i}}{\many{\rho_i}}}
                      {\Subst{e}{\many{\alpha_i}}{\many{\rho_i}}}
                      {\Subst{\tau}{\many{\alpha_i}}{\many{\rho_i}}}
                      $ when, for each $i$, $\Kinding{\Delta}{\rho_i}{\kappa_i}$.
\end{lemma}
\noindent The above lemma is sufficient to show the monomorphic instantiation case, by setting $\Delta = \varepsilon$ (the empty context). This lemma is a
key ingredient for the refinement link between polymorphic and monomorphic deep embeddings (See \autoref{s:mono-correctness}).

\subsection{Dynamic Semantics}
\label{sec:dynsem}
\begin{figure}
  \boxlabel{Value Semantics}
  \begin{grammar}
    \text{values} & v & \Coloneqq & \ell \alt \Unit \\ 
                  &   & \alt      & \FunVal{x}{e}                     & \text{(function values)} \\
                  &   & \alt      & \AbsFunVal{f}{\many{\tau}}        & \text{(abstract functions)} \\
                  &   & \alt      & \Cons{C}{v}                       & \text{(variant values)} \\
                  &   & \alt      & \RecordVal{\many{\FieldEq{f}{v}}} & \text{(records)} \\
                  &   & \alt      & a_v                               & \text{(abstract values)} \\
    \text{environments} & V & \Coloneqq & \many{\EnvBind{x}{v}} \\
    \text{abstract values} & a_v 
  \end{grammar}
  \begin{displaymath}
   \begin{array}{lclr}
    \AValSem{\cdot} & : & f \rightarrow (v \rightarrow v) & \text{(abstract function semantics)}
   \end{array}
  \end{displaymath}
  \boxlabel{Update Semantics}
  \begin{grammar}
    \text{u. sem. values} & u & \Coloneqq & \ell \alt \Unit \\ 
                            &   & \alt      & \FunVal{x}{e}                     & \text{(function values)} \\
                            &   & \alt      & \AbsFunVal{f}{\many{\tau}}        & \text{(abstract functions)} \\
                            &   & \alt      & \Cons{C}{u}                       & \text{(variant values)} \\
                            &   & \alt      & \RecordVal{\many{\FieldEq{f}{u}}} & \text{(records)} \\
                            &   & \alt      & a_u                               & \text{(abstract values)} \\
                            &   & \alt      & p                                 & \text{(pointers)} \\
    \text{environments}     & U & \Coloneqq & \many{\EnvBind{x}{u}}\\
    \text{pointers}         & p & & \multicolumn{2}{l}{$\text{sets of pointers}$\ \hspace{5ex} $r, w$} \\
    \text{abstract values}  & a_u & & 
    \multicolumn{2}{l}{\text{stores}\ \hspace{13ex} \ensuremath{\mu : p \nrightarrow u}}\\
  \end{grammar}
  \begin{displaymath}
   \begin{array}{lclr}
    \AUpdSem{\cdot} & : & f \rightarrow (u \times \mu \rightarrow u \times \mu) & \text{(abstract function semantics)}
   \end{array}
  \end{displaymath}
  \caption{Definitions for Value and Update Semantics}
  \label{fig:semdef}
\end{figure}
\autoref{fig:updvalsem} defines the big-step evaluation rules for the 
\emph{value} semantics of \CDSL. The relation
\mbox{$\ValSem{V}{e}{v}$} states that under
environment $V$, the expression $e$ evaluates to a resultant value
$v$. These values are documented in \autoref{fig:semdef}. In many
ways, the semantics is entirely typical of a purely functional
language, albeit with some care to handle abstract function calls
appropriately. This is intentional, since our goal is to automatically
produce a purely functional shallow embedding from this semantics.

As functions must be defined on the top level, our function values
$\FunVal{x}{e}$ consist only of an unevaluated expression, which is
evaluated when the function is applied. Abstract function values,
written $\AbsFunVal{f}{\many{\tau}}$, are instead passed more
indirectly, as a pair of the function name and a list of the types
used to instantiate any type variables.  When an abstract function
value $\AbsFunVal{f}{\many{\tau}}$ is applied, the user-supplied
semantics $\AValSem{f}$ are invoked, which is simply a function from
input value to output value.

The \emph{update} semantics, by contrast, is much more imperative. The
semantic rules can also be found in \autoref{fig:updvalsem}, with associated
definitions in
\autoref{fig:semdef}. This semantics is also an evaluation
semantics, written $\UpdSem{U}{e}{\mu}{u}{\mu'}$ in the style 
of~\citet{Pierce_02}. Values in the update semantics may now be 
\emph{pointers}, written $p$, to values in a mutable store or \emph{heap} $\mu$.  
This mutable store is modelled as a partial function from a pointer to 
an update semantics value. 

Most of the rules in \autoref{fig:updvalsem} only differ from the value
semantics in that they thread the store~$\mu$ through the
evaluation of the program. However, the key differences arise in the
treatment of records and of abstract types, which may now be
represented as \emph{boxed} structures, stored on the heap. In
particular, note that the rule $\rulename{UPut}_2$ destructively
updates the heap, instead of creating a new record value, and the
semantics of abstract functions $\AUpdSem{\cdot}$ may also modify the
heap.

\renewcommand{\rulename}[1]{\textsc{\scriptsize #1}}
\begin{figure*}[t]
\small
  \begin{inductive}{\ValSem{V}{e}{v}}
    \inferrule{(\EnvBind{x}{v}) \in V}{\ValSem{V}{x}{v}}{\rulename{VVar}} \quad\!
    \inferrule{ }{\ValSem{V}{\Unit}{\Unit}}{\rulename{V()}} \quad\!
    \inferrule{\FunDefn{f} = \FunDef{f}{\PolyTy{\many{\OfKind{\alpha_i}{\kappa_i}}}{\FunTy{\tau}{\rho}}}{x}{e}}
              {\ValSem{V}{\TyApp{f}{\many{\tau_i}}}{\FunVal{x}{\Subst{e}{\many{\alpha_i}}{\many{\tau_i}}}}}
              {\rulename{VFun}_C}\quad\!
    \inferrule{\FunDefn{f} = \AbsFunDef{f}{\PolyTy{\many{\OfKind{\alpha_i}{\kappa_i}}}{\FunTy{\tau}{\rho}}}}
              {\ValSem{V}{\TyApp{f}{\many{\tau_i}}}{\AbsFunVal{f}{\many{\tau_i}}}}
              {\rulename{VFun}_A}\\
    \inferrule{ }{\ValSem{V}{\ell}{\ell}}{\rulename{VLit}} \quad\!\!\!\!
    \inferrule{ \ValSem{V}{e}{\ell}}{\ValSem{V}{\Cast{t}{e}}{\ell}}{\rulename{VCast}} \quad\!\!\!\!
    \inferrule{\ValSem{V}{e_1}{\FunVal{x}{e}} \\\\ \ValSem{V}{e_2}{v'} \quad\!\! \ValSem{(\EnvBind{x}{v'})}{e}{v}}
              {\ValSem{V}{\App{e_1}{e_2}}{v}}
              {\rulename{VApp}_C}\quad\!\!\!\!\!
    \inferrule{\ValSem{V}{e_1}{\AbsFunVal{f}{\many{\tau}}} \\\\ \ValSem{V}{e_2}{v'} \quad\!\! \AbsValSem{f}{v'}{v}}
              {\ValSem{V}{\App{e_1}{e_2}}{v} }{\rulename{VApp}_A} \quad\!\!\!\!\!
    \inferrule{\text{for each $i$:}\ \ValSem{V}{e_i}{\ell_i}}{\ValSem{V}{\GenPrimOp{o}{\many{e_i}}}{\GenPrimOp{o}{\many{\ell_i}}}}{\rulename{VPrimOp}}\\
    \inferrule{\ValSem{V}{e_1}{v'} \\\\ \ValSem{\EnvBind{x}{v'},V}{e_2}{v}}{\ValSem{V}{\Let{x}{e_1}{e_2}}{v}}{\rulename{VLet}}\quad
    \inferrule{\ValSem{V}{e_1}{v'} \\\\ \ValSem{\EnvBind{x}{v'},V}{e_2}{v}}{\ValSem{V}{\LetBang{\many{y}}{x}{e_1}{e_2}}{v}}{\rulename{VLet}!}\quad
    \inferrule{\ValSem{V}{e}{v}}{\ValSem{V}{\Cons{C}{e}}{\Cons{C}{v}}}{\rulename{VCons}} \quad
    \inferrule{\ValSem{V}{e}{\Cons{C_k}{v}}}{\ValSem{V}{\Promote{\many{\Cons{C_i}{\tau_i}}}{e}}{\Cons{C_k}{v}}}\rulename{VProm}\\
    \inferrule{\ValSem{V}{e_1}{\Cons{C}{v'}} \\ \ValSem{\EnvBind{x}{v'}, V}{e_2}{v}}{\ValSem{V}{\Case{e_1}{\Cons{C}{\VarN{x}}}{e_2}{y}{e_3}}{v}}{\rulename{VCase}_1} \quad
    \inferrule{\ValSem{V}{e_1}{\Cons{B}{v'}} \\ \ConsN{B} \neq \ConsN{C} \\ \ValSem{\EnvBind{x}{(\Cons{B}{v'})}, V}{e_3}{v}}
              {\ValSem{V}{\Case{e_1}{\Cons{C}{\VarN{x}}}{e_2}{y}{e_3}}{v}}{\rulename{VCase}_2} \quad
    \inferrule{\ValSem{V}{e}{\Cons{C}{v}}}{\ValSem{V}{\Esac{e}}{v}}{\rulename{VEsac}} \\
    \inferrule{\text{for each $i$:}\ \ValSem{V}{e_i}{v_i}}
              {\ValSem{V}{\StructInit{\many{\FieldEq{f$_i$}{e_i}}}}{\RecordVal{\many{\FieldEq{f$_i$}{v_i}}}}}
              {\rulename{VStr}} \quad\!\!\!
    \inferrule{\ValSem{V}{e}{\RecordVal{\many{\FieldEq{f$_i$}{v_i}}}}}
              {\ValSem{V}{\Member{e}{f$_k$}}{v_k}}
              {\rulename{VMem}} \quad\!\!\!
    \inferrule{\ValSem{V}{e_1}{\RecordVal{\many{\FieldEq{f$_i$}{v_i}}}} 
              \\\\ \ValSem{\EnvBind{x}{\RecordVal{\many{\FieldEq{f$_i$}{v_i}}}}, \EnvBind{y}{v_k}, V}{e_2}{v} }
              {\ValSem{V}{\Take{x}{f$_k$}{y}{e_1}{e_2}}{v}}{\rulename{VTake}}\quad\!\!\!
    \inferrule{\ValSem{V}{e_1}{\RecordVal{\many{\FieldEq{f$_i$}{v_i}}}} 
              \quad \ValSem{V}{e_2}{v'_k}
              \\\\ \text{for each $i \neq k$:}\ v'_i = v_i}
              {\ValSem{V}{\Put{e_1}{f$_k$}{e_2}}{\RecordVal{\many{\FieldEq{f$_i$}{v'_i}}}}}{\rulename{VPut}}
  \end{inductive}
\vspace{0.5ex}
  \begin{inductive}{\UpdSem{U}{e}{\mu}{u}{\mu'}}
    \inferrule{\FunDefn{f} = \FunDef{f}{\PolyTy{\many{\OfKind{\alpha_i}{\kappa_i}}}{\FunTy{\tau}{\rho}}}{x}{e}}
              {\UpdSem{U}{\TyApp{f}{\many{\tau_i}}}{\mu}{\FunVal{x}{\Subst{e}{\many{\alpha_i}}{\many{\tau_i}}}}{\mu}}
              {\rulename{UFun}_C}\quad\!
    \inferrule{\FunDefn{f} = \AbsFunDef{f}{\PolyTy{\many{\OfKind{\alpha_i}{\kappa_i}}}{\FunTy{\tau}{\rho}}}}
              {\UpdSem{U}{\TyApp{f}{\many{\tau_i}}}{\mu}{\AbsFunVal{f}{\many{\tau_i}}}{\mu}}
              {\rulename{UFun}_A} \quad\!
    \inferrule{\UpdSem{U}{e_1}{\mu}{u'}{\mu_1} \\\\ \UpdSem{\EnvBind{x}{u'},U}{\mu_1}{e_2}{u}{\mu_2}}{\UpdSem{U}{\Let{x}{e_1}{e_2}}{\mu}{u}{\mu_2}}{\rulename{ULet}}\\
    \inferrule{\UpdSem{U}{e_1}{\mu}{\FunVal{x}{e}}{\mu_1} \\\\ \UpdSem{U}{e_2}{\mu_1}{u'}{\mu_2} \quad\!\! \UpdSem{(\EnvBind{x}{u'})}{e}{\mu_2}{u}{\mu_3}}
              {\UpdSem{U}{\App{e_1}{e_2}}{\mu}{u}{\mu_3}}
              {\rulename{UApp}_C}\quad\!
    \inferrule{\UpdSem{U}{e_1}{\mu}{\AbsFunVal{f}{\many{\tau_i}}}{\mu_1} \\\\ \UpdSem{U}{e_2}{\mu_1}{u'}{\mu_2} \quad\!\! \AbsUpdSem{f}{u'}{\mu_2}{u}{\mu_3}}
              {\UpdSem{U}{\App{e_1}{e_2}}{\mu}{u}{\mu_3} }{\rulename{UApp}_A} \quad\!
    \inferrule{\UpdSem{U}{e_1}{\mu}{u'}{\mu_1} \\\\ \UpdSem{\EnvBind{x}{u'},U}{\mu_1}{e_2}{u}{\mu_2}}{\UpdSem{U}{\LetBang{\many{y}}{x}{e_1}{e_2}}{\mu}{u}{\mu_2}}{\rulename{ULet!}}\\ 
    \inferrule{\UpdSem{U}{e}{\mu}{\Cons{C_k}{u}}{\mu'}}{\UpdSem{U}{\Promote{\many{\Cons{C_i}{\tau_i}}}{e}}{\mu}{\Cons{C_k}{u}}{\mu'}}\rulename{UProm}\quad\!\!
    \inferrule{\UpdSem{U}{e}{\mu}{\Cons{C}{u}}{\mu'}}{\UpdSem{U}{\Esac{e}}{\mu}{u}{\mu'}}{\rulename{UEsac}} \quad\!\!
    \inferrule{\UpdSemA{U}{\many{e_i}}{\mu}{\many{u_i}}{\mu'}}
              {\UpdSem{U}{\StructInit{\many{\FieldEq{f$_i$}{e_i}}}}{\mu}{\RecordVal{\many{\FieldEq{f$_i$}{u_i}}}}{\mu'}}
              {\rulename{UStr}} \quad\!\!\!
    \inferrule{\UpdSem{U}{e}{\mu}{\RecordVal{\many{\FieldEq{f$_i$}{u_i}}}}{\mu'}}
              {\UpdSem{U}{\Member{e}{f$_k$}}{\mu}{u_k}{\mu'}}
              {\rulename{UMem}_1} 
              \\
    \inferrule{\UpdSem{U}{e_1}{\mu}{\Cons{C}{u'}}{\mu_1} \\\\ \UpdSem{\EnvBind{x}{u'}, U}{\mu_1}{e_2}{u}{\mu_2}}
              {\UpdSem{U}{\Case{e_1}{\Cons{C}{\VarN{x}}}{e_2}{y}{e_3}}{\mu}{u}{\mu_2}}{\rulename{UCase}_1} \quad\!\!\!
    \inferrule{\UpdSem{U}{e_1}{\mu}{\Cons{B}{u'}}{\mu_1} \\ \ConsN{B} \neq \ConsN{C} \\\\ \UpdSem{\EnvBind{x}{(\Cons{B}{u'})}, U}{e_3}{\mu_1}{u}{\mu_2}}
              {\UpdSem{U}{\Case{e_1}{\Cons{C}{\VarN{x}}}{e_2}{y}{e_3}}{\mu}{u}{\mu_2}}{\rulename{UCase}_2} \quad\!\!\!
    \inferrule{\UpdSem{U}{e}{\mu}{p}{\mu'} \\\\ \mu'(p) = {\RecordVal{\many{\FieldEq{f$_i$}{u_i}}}}}
              {\UpdSem{U}{\Member{e}{f$_k$}}{\mu}{u_k}{\mu'}}
              {\rulename{UMem}_2}
    \end{inductive}
    \begin{center}
        \begin{inductive0}
    \inferrule{(\EnvBind{x}{u}) \in U}{\UpdSem{U}{x}{\mu}{u}{\mu}}{\rulename{UVar}} \quad\!
          \inferrule{\UpdSem{U}{e_1}{\mu}{\RecordVal{\many{\FieldEq{f$_i$}{u_i}}}}{\mu_1} 
                \\\\ \UpdSem{\EnvBind{x}{\RecordVal{\many{\FieldEq{f$_i$}{u_i}}}}, \EnvBind{y}{u_k}, U}{e_2}{\mu_1}{u}{\mu_2} }
                    {\UpdSem{U}{\Take{x}{f$_k$}{y}{e_1}{e_2}}{\mu}{u}{\mu_2}}{\rulename{UTake}_1}\quad\!\!\!
          \inferrule{\UpdSem{U}{e_1}{\mu}{\RecordVal{\many{\FieldEq{f$_i$}{u_i}}}}{\mu_1} 
                \\\\ \UpdSem{U}{e_2}{\mu_1}{u'_k}{\mu_2}
                \quad \text{for each $i \neq k$:}\ u'_i = u_i}
                    {\UpdSem{U}{\Put{e_1}{f$_k$}{e_2}}{\mu}{\RecordVal{\many{\FieldEq{f$_i$}{u'_i}}}}{\mu_2}}{\rulename{UPut}_1} \\[1.5em]
    \inferrule{ \UpdSem{U}{e}{\mu}{\ell}{\mu'}}{\UpdSem{U}{\Cast{t}{e}}{\mu}{\ell}{\mu'}}{\rulename{UCast}} \quad\!\!\!\!
          \inferrule{\UpdSem{U}{e_1}{\mu}{p}{\mu_1} 
                \\   \mu_1(p) = {\RecordVal{\many{\FieldEq{f$_i$}{u_i}}}}
                \\\\ \UpdSem{\EnvBind{x}{p}, \EnvBind{y}{u_k}, U}{e_2}{\mu_1}{u}{\mu_2} }
                    {\UpdSem{U}{\Take{x}{f$_k$}{y}{e_1}{e_2}}{\mu}{u}{\mu_2}}{\rulename{UTake}_2}\quad\!\!\!
          \inferrule{\UpdSem{U}{e_1}{\mu}{p}{\mu_1}
               \quad \UpdSem{U}{e_2}{\mu_1}{u'_k}{\mu_2}
               \\\\  \mu_2(p) = {\RecordVal{\many{\FieldEq{f$_i$}{u_i}}}}
               \\    \text{for each $i \neq k$:}\ u'_i = u_i}
                    {\UpdSem{U}{\Put{e_1}{f$_k$}{e_2}}{\mu}{p}{\mu_2(p \coloneqq \RecordVal{\many{\FieldEq{f$_i$}{u'_i}}} )}}{\rulename{UPut}_2}
        \end{inductive0}
    \end{center}
  \caption{\CDSL Value and Update Semantics (some straightforward rules omitted for brevity)}
  \label{fig:updvalsem}
\end{figure*}

\subsubsection{Update-Value Refinement and Type Preservation}
\label{sec:updvalrel}
\begin{figure*}
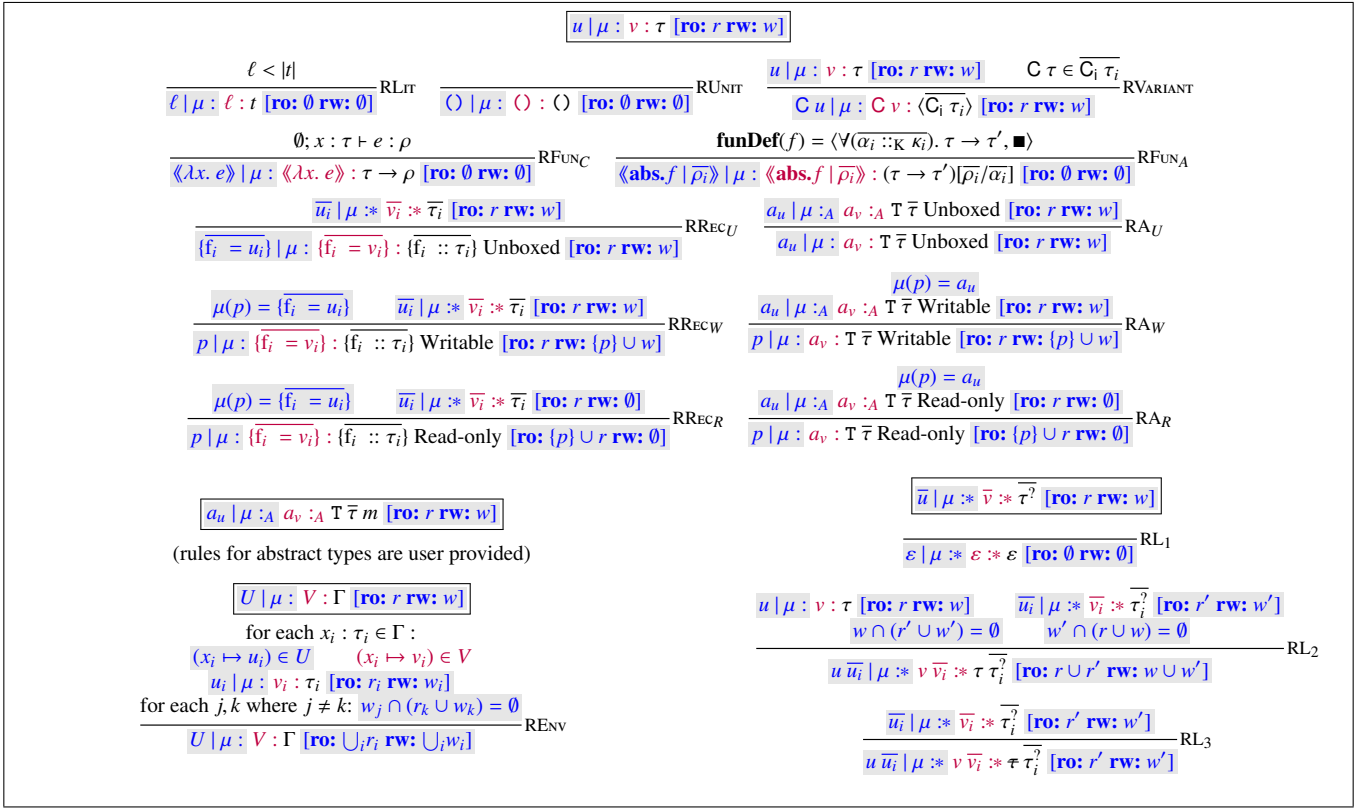

  \small
  \begin{inductive}{\VTR{u}{\mu}{v}{\tau}{r}{w}}
  \inferrule{ \ell < |t|}
            {\VTR{\ell}{\mu}{\ell}{t}{\emptyset}{\emptyset}}
            {\rulename{RLit}}\quad
  \inferrule{ }{\VTR{\Unit}{\mu}{\Unit}{\Unit}{\emptyset}{\emptyset}}{\rulename{RUnit}}\quad
  \inferrule{ \VTR{u}{\mu}{v}{\tau}{r}{w} \\ 
              \Cons{C}{\tau} \in \many{\Cons{C_i}{\tau_i}}
             }{\VTR{\Cons{C}{u}}{\mu}{\Cons{C}{v}}{\VariantTy{\many{\Cons{C_i}{\tau_i}}}}{r}{w}}{\rulename{RVariant}}\\
  \inferrule{\Typing{\emptyset}{\ofType{x}{\tau}}{e}{\rho}}
            {\VTR{\FunVal{x}{e}}{\mu}{\FunVal{x}{e}}{\FunTy{\tau}{\rho}}{\emptyset}{\emptyset}}{\rulename{RFun}_C}\quad
  \inferrule{\AbsFunTyEnv{f}{\PolyTy{\many{\OfKind{\alpha_i}{\kappa_i}}}{\tau \rightarrow \tau'}}}
            {\VTR{\AbsFunVal{f}{\many{\rho_i}}}{\mu}{\AbsFunVal{f}{\many{\rho_i}}}{\Subst{(\FunTy{\tau}{\tau'})}{\many{\alpha_i}}{\many{\rho_i}}}{\emptyset}{\emptyset}}{\rulename{RFun}_A}\\
  \inferrule{\VTRA{\many{u_i}}{\mu}{\many{v_i}}{\many{\tau_i}}{r}{w}}
            {\VTR{\RecordVal{\many{\FieldEq{f$_i$}{u_i}}}}{\mu}{\RecordVal{\many{\FieldEq{f$_i$}{v_i}}}}{\RecordTy{\many{\FieldTy{f$_i$}{\tau_i}}}{\Unboxed}}{r}{w}}{\rulename{RRec}_U}\quad
  \inferrule{\VTRAbs{a_u}{\mu}{a_v}{\AbsTy{T}{\many{\tau}}{\Unboxed}}{r}{w}}{\VTR{a_u}{\mu}{a_v}{\AbsTy{T}{\many{\tau}}{\Unboxed}}{r}{w}}{\rulename{RA}_U} \\
  \inferrule{\HiBlue{\mu(p) = \RecordVal{\many{\FieldEq{f$_i$}{u_i}}}} \\
             \VTRA{\many{u_i}}{\mu}{\many{v_i}}{\many{\tau_i}}{r}{w} }
            {\VTR{p}{\mu}{\RecordVal{\many{\FieldEq{f$_i$}{v_i}}}}{\RecordTy{\many{\FieldTy{f$_i$}{\tau_i}}}{\Writable}}{r}{\{p\} \cup w}}{\rulename{RRec}_W}\quad
  \inferrule{\HiBlue{\mu(p) = a_u}\\\\\VTRAbs{a_u}{\mu}{a_v}{\AbsTy{T}{\many{\tau}}{\Writable}}{r}{w}} 
            {\VTR{p}{\mu}{a_v}{\AbsTy{T}{\many{\tau}}{\Writable}}{r}{\{p\} \cup w}}{\rulename{RA}_W} \\
  \inferrule{\HiBlue{\mu(p) = \RecordVal{\many{\FieldEq{f$_i$}{u_i}}}} \\
             \VTRA{\many{u_i}}{\mu}{\many{v_i}}{\many{\tau_i}}{r}{\emptyset} }
            {\VTR{p}{\mu}{\RecordVal{\many{\FieldEq{f$_i$}{v_i}}}}{\RecordTy{\many{\FieldTy{f$_i$}{\tau_i}}}{\ReadOnly}}{\{p\} \cup r}{\emptyset}}{\rulename{RRec}_R} \quad
  \inferrule{\HiBlue{\mu(p) = a_u}\\\\\VTRAbs{a_u}{\mu}{a_v}{\AbsTy{T}{\many{\tau}}{\ReadOnly}}{r}{\emptyset}} 
            {\VTR{p}{\mu}{a_v}{\AbsTy{T}{\many{\tau}}{\ReadOnly}}{\{p\} \cup r}{\emptyset}}{\rulename{RA}_R} 
  \end{inductive}
  \begin{sidebyside}
  \begin{inductive}{\VTRAbs{a_u}{\mu}{a_v}{\AbsTy{T}{\many{\tau}}{m}}{r}{w}}
    \text{(rules for abstract types are user provided)}
  \end{inductive}
  \begin{inductive}{\VTR{U}{\mu}{V}{\Gamma}{r}{w}}
  \inferrule{\text{for each $ \ofType{x_i}{\tau_i} \in \Gamma$}: \\\\
              \HiBlue{(\EnvBind{x_i}{u_i}) \in U} \\ \HiPurple{(\EnvBind{x_i}{v_i}) \in V} \\\\
              \VTR{u_i}{\mu}{v_i}{\tau_i}{r_i}{w_i} \\\\
             \text{for each $j, k$ where $j \neq k$:}\ 
             \HiBlue{w_j \cap (r_k \cup w_k) = \emptyset}}
             {\VTR{U}{\mu}{V}{\Gamma}{\textstyle{\bigcup}_i r_i}{\textstyle{\bigcup}_i w_i}}{\rulename{REnv}}
  \end{inductive}
  & 
  \vspace{-1em}
  \begin{inductive}{\VTRA{\many{u}}{\mu}{\many{v}}{\many{\perhaps{\tau}}}{r}{w}}
    \inferrule{ }{\VTRA{\varepsilon}{\mu}{\varepsilon}{\varepsilon}{\emptyset}{\emptyset}}{\rulename{RL}_1} \\[0.4em]
    \inferrule{ \VTR{u}{\mu}{v}{\tau}{r}{w}
            \\  \VTRA{\many{u_i}}{\mu}{\many{v_i}}{\many{\perhaps{\tau_i}}}{r'}{w'}
            \\\\ \HiBlue{w \cap (r' \cup w') = \emptyset} \\
               \HiBlue{w' \cap (r \cup w) = \emptyset}}
              {\VTRA{u\ \many{u_i}}{\mu}{v\ \many{v_i}}{\tau\ \many{\perhaps{\tau_i}}}{r \cup r'}{w \cup w'}}{\rulename{RL}_2} \\[0.4em]
    \inferrule{\VTRA{\many{u_i}}{\mu}{\many{v_i}}{\many{\perhaps{\tau_i}}}{r'}{w'}}
              {\VTRA{u\ \many{u_i}}{\mu}{v\ \many{v_i}}{\taken{\tau}\ \many{\perhaps{\tau_i}}}{r'}{w'}}{\rulename{RL}_3} 
  \end{inductive} 
  \end{sidebyside}
\caption{Value Typing and Refinement.  For value typing rules, erase \HiBlue{\text{this}} text for value semantics, and \HiPurple{this} text for update semantics.}
\label{fig:valtypref}
\end{figure*}

In order to show that the update semantics is a refinement of the value semantics, we must exploit the information given to us by \CDSL's linear type system. 
A typical refinement approach to relate the two semantics would be to define a correspondence relation between update semantics states and value semantics values, 
and show that an update semantics evaluation implies a corresponding value semantics evaluation. However, such a statement is not true if aliasing exists, as a destructive
update (from, say, $\PUT$) would result in multiple values being changed in the update semantics but not necessarily in the value semantics. As our type system forbids
aliasing of writable references, we must include this information in our correspondence relation. Written as $u\ |\ \mu : v : \tau\ [\textbf{ro:}\ r\ \textbf{rw:}\ w]$, this relation
states that the update semantics value $u$ with store $\mu$ corresponds to the value semantics value $v$, which both have the type $\tau$. The sets $r$ and $w$ contain
all pointers accessible from the value $u$ that are read-only and writable respectively. We use this to encode the uniqueness property ensured by linear types as explicit
non-aliasing constraints in the rules for the correspondence relation, which are given in \autoref{fig:valtypref}. Read-only pointers may alias other read-only pointers, but
writable pointers do not alias any other pointer, whether read-only or writable. 

Because our correspondence relation includes types, it naturally implies a value typing relation for both value semantics (written $v : \tau$) and update semantics (written
$u\ |\ \mu : \tau\ [\textbf{ro:}\ r\ \textbf{rw:}\ w]$ ). In fact, the rules for both relations can be derived from the rules in \autoref{fig:valtypref} simply by
erasing either the value semantics parts (highlighted like \HiPurple{this}) or the update semantics parts (highlighted like \HiBlue{\text{this}}). 
As we ultimately prove preservation for this correspondence relation across evaluation, this same erasure strategy can be applied to our proofs to produce a type preservation
proof for either semantics.  

\paragraph{Formalising uniqueness}
With this correspondence relation, we can prove our intuitions about linear types. For example, the following lemma, which shows that we do not discard any unique
writable reference via weakening, makes use of the fact that a value is only given a discardable type when it contains no writable pointers.
\begin{lemma}[Weakening respects environment typing]$\ $ 

\noindent If $\VTR{U}{\mu}{V}{\Gamma}{r}{w}$ and $\Weakening{}{\Gamma}{\Gamma'}$ then there exists \HiBlue{r' \subseteq r} such that 
$\VTR{U}{\mu}{V}{\Gamma'}{r'}{w}$. 
\end{lemma}

\noindent We also prove a similar lemma about our context splitting judgement, which uses the fact that a value is only given a shareable type when it contains no writable pointers to conclude that the two output contexts give access to non-aliasing sets of writable pointers.
\begin{lemma}[Splitting respects environment typing]$\ $

\noindent If $\VTR{U}{\mu}{V}{\Gamma}{r}{w}$ and $\Contraction{}{\Gamma}{\Gamma_1}{\Gamma_2}$ then there exists \HiBlue{r_1, r_2} and \HiBlue{w_1, w_2} where \HiBlue{r = r_1 \cup r_2} and \HiBlue{w = w_1 \cup w_2}, such that $\VTR{U}{\mu}{V}{\Gamma_1}{r_1}{w_1}$ and $\VTR{U}{\mu}{V}{\Gamma_2}{r_2}{w_2}$ and \HiBlue{w_1 \cap w_2 = \emptyset}.
\end{lemma}

\noindent In addition, we prove our main intuition about $\BangF{\cdot}$, necessary for showing refinement for $\LET!$ expressions.
\begin{lemma}[$\BangF{\cdot}$ makes writable read-only]$\ $

\noindent If $\VTR{u}{\mu}{v}{\tau}{r}{w}$ then $\VTR{u}{\mu}{v}{\BangF{\tau}}{r \cup w}{\emptyset}$
\end{lemma}

\newcommand{\Frame}[4]{#1\ |\ #2\ \textbf{frame}\ #3\ |\ #4}
\paragraph{Dealing with mutable state} We define a \emph{framing} relation which specifies exactly how evaluation may affect the mutable store $\mu$. Given an input
set of writable pointers $w_i$, an input store $\mu_i$, an output set of pointers $w_o$ and an output store $\mu_o$, the relation, written $\Frame{w_i}{\mu_i}{w_o}{\mu_o}$, 
ensures three properties for any pointer $p$:
\begin{description}
  \item[Inertia] If $p \notin w_i \cup w_o$, then $\mu_i(p) = \mu_o(p)$.
  \item[Leak freedom] If $p \in w_i$ and $p \notin w_o$, then $\mu_o(p) = \bot$.
  \item[Fresh allocation] If $p \notin w_i$ and $p \in w_o$, then $\mu_i(p) = \bot$.
\end{description}
Framing implies that our correspondence relation, for both values and environments, is unaffected by unrelated store updates:

\begin{lemma}[Unrelated updates] Assume two unrelated pointer sets \HiBlue{w \cap w_1 = \emptyset} and that \HiBlue{\Frame{w_1}{\mu}{w_2}{\mu'}}, then

\begin{itemize}
\item If $\VTR{u}{\mu}{v}{\tau}{r}{w}$  
then $\VTR{u}{\mu'}{v}{\tau}{r}{w}$ and \HiBlue{w \cap w_2 = \emptyset}.
\item If $\VTR{U}{\mu}{V}{\Gamma}{r}{w}$ 
then $\VTR{U}{\mu'}{V}{\Gamma}{r}{w}$ and \HiBlue{w \cap w_2 = \emptyset}.
\end{itemize}
\end{lemma}

\paragraph{Refinement and preservation}
With the above lemmas and definitions, we are able to prove refinement between the value and the update semantics. This of course requires us to assume the same for
the semantics given to abstract functions, $\AValSem{\cdot}$ and $\AUpdSem{\cdot}$.

\begin{assumption} Let $f$ be an abstract function with type signature $f :: \PolyTy{\many{\OfKind{\alpha_i}{\kappa_i}}}{\tau \rightarrow \tau'}$, and $\many{\rho_i}$ be an
instantiation of the type variables $\many{\alpha_i}$ such that for each $i$, $\Kinding{}{\rho_i}{\kappa_i}$. Let $\HiBlue{u}$ and $\HiPurple{v}$ be update- and value-semantics values such that $\VTR{u}{\mu}{v}{\Subst{\tau}{\many{\alpha_i}}{\many{\rho_i}}}{r}{w}$. The user-supplied meaning of $f$ in each semantics gives $\HiPurple{\AbsValSem{f}{v}{v'}}$ and $\HiBlue{\AbsUpdSem{f}{u}{\mu}{u'}{\mu'}}$. 
Then, there exists \HiBlue{r' \subseteq r} and \HiBlue{w'} such that $\VTR{u'}{\mu'}{v'}{\Subst{\tau'}{\many{\alpha_i}}{\many{\rho_i}}}{r}{w}$ and \HiBlue{\Frame{w}{\mu}{w'}{\mu'}}.
\end{assumption}

\noindent We first prove that the correspondence relation is preserved when both semantics evaluate from corresponding environments. By erasing one semantics, this becomes
a type preservation theorem for the other. Due to space constraints, we omit the details of the proof in this paper, but the full proof is available in our Isabelle/HOL
formalisation.
\begin{theorem}[Preservation of types and correspondence]$\ $
\label{thm:preservation}
\noindent If $\Typing{\varepsilon}{\Gamma}{e}{\tau}$ and $\VTR{U}{\mu}{V}{\Gamma}{r}{w}$ and $\HiPurple{\ValSem{V}{e}{v}}$ and $\HiBlue{\UpdSem{U}{e}{\mu}{u}{\mu'}}$, 
then there exists \HiBlue{r' \subseteq r} and \HiBlue{w'} such that $\VTR{u}{\mu'}{v}{\tau}{r'}{w'}$ and \HiBlue{\Frame{w}{\mu}{w'}{\mu'}}.
\end{theorem}

\noindent In order to prove refinement, we must show that every evaluation on the concrete update semantics has a corresponding evaluation in the abstract value semantics. While \autoref{thm:preservation} already gets us most of the way there, we still need to prove that the value semantics can evaluate whenever the update semantics does.
\begin{lemma}[Upward-propagation of evaluation]$\ $

\noindent If $\Typing{\varepsilon}{\Gamma}{e}{\tau}$ and $\VTRN{U}{\mu}{V}{\Gamma}{r}{w}$ and $\UpdSem{U}{e}{\mu}{u}{\mu'}$, then there exists a $v$ such that $\ValSem{V}{e}{v}$
\end{lemma}

\noindent Composing this lemma and \autoref{thm:preservation}, we can now easily prove our desired refinement statement.
\begin{theorem}[Value $\Rightarrow$ Update refinement]
\label{thm:updvalrefinement}

\noindent If $\Typing{\varepsilon}{\Gamma}{e}{\tau}$ and $\VTRN{U}{\mu}{V}{\Gamma}{r}{w}$ and $\UpdSem{U}{e}{\mu}{u}{\mu'}$, then there exists a value $v$ and pointer sets $r' \subseteq r$ and  $w'$ such that $\ValSem{V}{e}{v}$, and $\VTRN{u}{\mu'}{v}{\tau}{r'}{w'}$ and $\Frame{w}{\mu}{w'}{\mu'}$.
\end{theorem}

\section{Verification}\label{s:verification}
With the formal semantics of \CDSL available, this section describes each of
the proof steps that make up the compiler certificate, depicted in
\autoref{fig:refinement} in \autoref{s:overview}.

\subsection{Top-Level Theorem}\label{s:toplevel}
We start by describing the top-level theorem that forms the program certificate, emitted by the compiler. Recall that for a well-typed
\CDSL program, the compiler produces C code, a shallow
embedding in Isabelle/HOL, and a refinement proof between them.

We say a C program correctly implements its \CDSL shallow embedding if the following holds:
\begin{inparaenum}[(i)]
\item the C program terminates with defined execution; and 
\item if the initial C state and \CDSL store are related, and the input values of the
programs are related, then their output values are related.
\end{inparaenum}

This means, the compiler correctness theorem states that a \emph{value
relation} is preserved. This relation is concrete and can be inspected.
In~\autoref{sec:updvalrel}, we introduced a value typing relation between
update semantics and value semantics. At each other refinement stage in the
following sections, we will introduce a further relation between values of
the two respective programs. By composing these value relations, we get the
value relation $\mathcal{V}$ between the result $v_m$ of the C program $p_m$
and the shallow embedding $s$ by going through the intermediate update
semantics value $u$ and value semantics result $v$. Note that the relation
in~\autoref{sec:updvalrel} also depends on a \CDSL store $\mu$. The C state
and \CDSL store are related using the \emph{state relation} \srel, defined in
detail in \autoref{s:c-to-deep}.

Let $\lambda e. \ \monoexpr \ \rename \ e$ and 
  $\lambda v. \ \monoval \ \rename \ v$ (defined in~\autoref{s:mono})
  be two functions
   that monomorphise expressions and (function) values, respectively,
   using a rename function \rename provided by the compiler.
Further, let 
   $R$ be a state relation, 
   $s$ a shallow embedding, 
   $e$ a monomorphic deep embedding, 
   $p_m$  a C program, $\mu$ a \CDSL store and $\sigma$ a  C state. 
    Then we define 
$\correspond$  as follows: \\
     If ($\exists r\;w.\;\VTRN{U}{\mu}{V}{\Gamma}{r}{w})$ and $(\mu, \sigma) \in R$, 
     then $p_m$ successfully terminates starting at $\sigma$; and 
     after executing $p_m$, for any resulting value $v_m$ and state $\sigma^\prime$, there exist $\mu^\prime$, $u$, and $v$ such that:
$$(\mu^\prime, \sigma^\prime) \in \srel \ \wedge 
    \UpdSem{U}{e}{\mu}{u}{\mu^\prime}\wedge \
    \ValSem{V}{e}{\monoval \ \rename \ v} \wedge \
       \mathcal{V} \  r \ \mu^\prime \ v_m\ u\ v \ s $$
%


\begin{theorem}
Given a \CDSL function~$f$ that takes $x$ of type $\tau$ as input, 
let~$p_m$ be its generated C code, 
$s$ its shallow embedding, and 
$e$ its deep embedding. 
Let~$v_m$ be an argument of~$p_m$, and~$u$ and $v$ be the update and value semantics arguments, of appropriate type, for~$f$. If $r$ is injective, then
\[
\begin{array}{l}
    \forall \mu\ \sigma.\ \mathcal{V} \ \rename \ \mu  \ v_m\ u\ v \ s \  \longrightarrow \\
   \quad\quad\correspond \ \rename \ \srel
     \  (s \; v_s) \ (\monoexpr \ \rename \ e) \ (p_m v_m)\ 
       U\  V\  \Gamma \ \mu \ \sigma
\end{array}
\]
\text {where  $U = (x \mapsto u)$, $V= (x \mapsto \ v)$, and $\Gamma = (x \mapsto \tau)$.}
\end{theorem}

\noindent
This top-level refinement theorem additionally assumes that abstract functions
in the program adhere to their specification and that their behaviour
remains the same when they are monomorphised.

Intuitively, this theorem states that for related input values, all programs in the refinement chain evaluate to related output values. This can of course be used to deduce that there exist intermediate programs through which the C code and its shallow embedding are directly related. The proof engineer does not need to care what those intermediate programs are.

%
%
%
%
%
%
%

\subsection{Well-typedness}\label{s:typetree}

Before we present each refinement step, we briefly describe the well-typing
theorems that are used in these steps.

The \CDSL compiler proves, via an automated Isabelle/HOL tactic, that the
monomorphic deep embedding of the input program is well-typed. Specifically,
the compiler defines $\FunDefn{\cdot}$ in Isabelle/HOL and proves that each
\CDSL function~$f\ \VarN{x} = e$ is well-typed in accordance with its type as
given by $\FunDefn{\cdot}$. Polymorphic well-typing is derived generically in
the monomorphisation proof in \autoref{s:mono-correctness}.

\begin{theorem}[Typing]
Let~$f$ be a (monomorphic) \CDSL function, where \mbox{$\FunDefn{f} = \FunDef{f}{\FunTy{\tau}{\tau'}}{x}{e}$}.
Then $\Typing{\varepsilon}{x : \tau}{e}{\tau'}$.
\end{theorem}

\noindent Because, as we will see in \autoref{s:c-to-deep}, proving
refinement requires access to the typing judgements for program
sub-expressions and not just for the top level, the \CDSL compiler also
instructs Isabelle to store all of the intermediate typing judgements
established during type checking. These theorems are stored in a tree
structure, isomorphic to the type derivation tree for the \CDSL program. Each
node is a typing theorem for a sub-expression of the program.

\subsection{From C to \CDSL Monomorphic Deep Embedding}\label{s:c-to-deep}

This section describes the first three transformations from
\autoref{fig:refinement} in \autoref{s:overview}. In the first step, the C
code is converted to Simpl~\citep{schirmer:phd} by the C-to-Isabelle
parser~\citep{Tuch_KN_07}, used in the seL4
project~\citep{Klein_EHACDEEKNSTW_09}. This step is kept as simple as
possible and makes no effort to abstract from the details of C.


\newcommand{\state}{\mathit{state}}
The second step in \autoref{fig:refinement}, which is the first link in the formal refinement chain, applies a modified version of the AutoCorres tool to produce a \emph{monadic} shallow embedding of the C code
semantics, and additionally proves that the Simpl C semantics is
a refinement of the monadic shallow embedding. We modify AutoCorres to make
its output more predictable by switching off its control-flow
simplification and forcing it to always output the shallow embedding in
the \emph{nondeterministic state monad} of~\citet{Cock_KS_08}.
In this monad, computation is represented by functions of type
$\state \Rightarrow (\alpha \mathrel{\times} \state) \;\set \mathrel{\times} \bool$. Here
$\state$ is the global state of the C program, including
global variables, while $\alpha$ is the return-type of the computation.
A computation takes as input the global state and returns a set, \results, of pairs with new state and result value. Additionally the computation
returns a boolean, \failed, indicating whether it failed (e.g. whether there was undefined behaviour). 

\newcommand{\word}{\mathsf{word}}
While AutoCorres was designed to facilitate manual reasoning about C code,
here we use it as the foundation for automatically proving correspondence to
the \CDSL input program. One of the main benefits AutoCorres gives us is a
\emph{typed} memory model. Specifically, the $\state$ of the AutoCorres monadic
representation contains a set of \emph{typed heaps}, each of type~$\mathsf{32}\  \word \Rightarrow \alpha$, 
one for each type~$\alpha$ used on the heap in the C input program. 

Proving that the
AutoCorres-generated monadic embedding never fails implies that the C
code is type- and memory-safe, and is free of undefined behaviour~\citep{Greenaway_LAK_14}. We prove
non-failure as a side-condition of the refinement statement from the
AutoCorres shallow embedding to the \CDSL monomorphic deep embedding in its
update semantics, essentially using \CDSL's type system to guarantee C memory
safety during execution.


This refinement proof is the third step in \autoref{fig:refinement}.
To phrase the refinement statement we first define how deeply-embedded 
\CDSL values and types relate to their corresponding monadic shallowly-embedded C values.
The value-mapping is captured by
the \emph{value relation}~$\valrel$, defined in Isabelle/HOL automatically
by the
\CDSL compiler using ad hoc
overloading. $\valrel$ is defined separately for each \CDSL program because the
types used in the shallow C embedding depend on those used in
the C program as, e.g., C structs are represented directly as
Isabelle/HOL records. 


The type relation $\typerel$ is used to determine, for a \CDSL value~$v$ of
type $\tau$, which typed heap in the state of the monadic shallow 
embedding~$v$ should appear in. As with $\valrel$ it is defined automatically
for each \CDSL program.

Given $\valrel$ and~$\typerel$ for a particular \CDSL program, the
\emph{state relation}~$\srel$ defines the correspondence between the store~$\mu$
over which the \CDSL update semantics operates, 
and the state~$\sigma$ of the monadic shallow embedding. 

\begin{definition}[Monad-to-Update State Relation]
$(\mu,\sigma) \in \srel$ if and only if: for all pointers~$p$
in the domain of $\mu$, there exists a
value~$v$ in the appropriate heap of~$\sigma$ (as defined by $\typerel$) at
location~$p$, such that $\valrel\  \mu(p)\  v$ holds.
\end{definition}

\noindent 
With $\srel$ and $\valrel$,
we define refinement generically between 
a monadic computation~$p_m$
and a \CDSL expression~$e$, evaluated under the update semantics. 
We denote the 
refinement predicate \corres.
Because $\srel$ changes for each \CDSL program, we parameterise
\corres by an arbitrary state relation~$R$. It is parameterised also by 
the typing context~$\Gamma$ and the environment~$U$, 
as well as by the initial update semantics store~$\mu$ and monadic shallow embedding
state~$\sigma$.

\begin{definition}{Monad-to-Update Correspondence}
\[
\begin{array}{l}
\corres \;R\; e\; p_m\;U\; \Gamma\; \mu\; \sigma = \\
\quad (\exists r \; w.\; U\ |\ \mu : \Gamma\ [\textbf{ro:}\ r\ \textbf{rw:}\ w])\longrightarrow \\
\quad \;\;     (\mu,\sigma) \in R \longrightarrow \\
\quad\quad   \;(\neg\ \failed \ (p_m \ \sigma)\; \wedge \\
\quad\quad   \;(\forall v_m \;\sigma^\prime. \ (v_m,\sigma^\prime) \in \results\ (p_m\;\sigma) \longrightarrow \\
\quad\quad\quad    (\exists \mu^\prime \ u. \ \UpdSem{U}{e}{\mu}{u}{\mu^\prime} \wedge 
      (\mu^\prime,\sigma^\prime) \in R \wedge \valrel\ u\ v_m)))
\end{array}
\]
\end{definition}

\noindent
The definition states that if the state relation~$R$ holds initially,
then the monadic computation~$p_m$ cannot fail and, moreover,
for all executions of~$p_m$ there must exist
a corresponding execution under the update semantics of the expression~$e$
such that the final states are related by~$R$ and $\valrel$ holds between
their results.
AutoCorres proves automatically that: $\neg \ \failed \ (p_m\  \sigma) \longrightarrow \results \ (p_m\ \sigma) \neq \emptyset$.

\paragraph{Refinement Proof}
The refinement proof is automatic in Isabelle, driven by a set of syntax-directed
\corres rules, one for each \cdsl construct. The proof procedure makes use
of the fact that the \cdsl term is in A-normal form to reduce the number of
cases that need to be considered and to simplify the higher-order unification
problems that some of the proof rules pose to Isabelle. 

This refinement theorem does not need an explicit formal assumption of
well-typedness of the \cdsl program. The proof tactic will simply fail for
programs that are not well-typed.

\autoref{fig:corres-rules} depicts two \corres rules, one for expressions~$x$
that are variables and the other for~$\Let{x}{a}{b}$.
These correspond respectively to the two basic monadic operations~\mreturn,
which yields values, and $>>=$, for sequencing computations.

Observe that the rule \textsc{Corres-Let} is \emph{compositional}: to prove
that $\Let{x}{a}{b}$ corresponds to $a' >>= b'$ the rule involves proving
that (1)~$a$ corresponds to $a'$ and (2)~that~$b$ corresponds to~$b'$ when each are
executed over corresponding results~$v_u$ and~$v_m$ (e.g. as yielded by
$a$ and $a'$ respectively). This compositionality significantly simplifies the automation of the correspondence proof.
The typing assumptions of \textsc{Corres-Let} are discharged by appealing to the type theorem tree generated by the compiler (see \autoref{s:typetree}). 

 \begin{figure} 
  \begin{tabular}{c} 
  \inferrule{ (\EnvBind{x}{u}) \in U  \\ \valrel\ u \ v_m}
            {\corresargs{x}{(\mreturn \  v_m)}}\; \rulename{Corres-Var}\\ \\ 

  \inferrule{\Contraction{}{\Gamma}{\Gamma_1}{\Gamma_2} \quad 
             \Typing{\varepsilon}{\Gamma_1}{a}{\tau} \quad
             \corresargsE{a}{a'}{U}{\Gamma_1} \\
             (\forall v_u \ v_m\ \mu^\prime\ \sigma^\prime.\ \valrel\ v_u v_m 
              \longrightarrow \\ 
              \quad\quad \corres \ R \ {b}\ (b' \; v_m)\
                                 (x \mapsto v_u,U) \ 
                                 (x : \tau,\Gamma_2) \ {\mu^\prime}
                                 \sigma^\prime
             )} 
            {\corresargs{(\Let{x}{a}{b})}{(a' >>= b')}} \!\rulename{Corres-Let} \\ 
  \end{tabular} \caption{Two example \corres rules}
  \label{fig:corres-rules}
\end{figure}

The rules for some of the other constructs, such as $\TAKE$, $\PUT$, and $\mathbf{case}$, 
contain non-trivial assumptions about \srel and about the types used 
in the program. Once a  program and its \srel are fixed, 
a set of simpler 
rules is automatically generated by \emph{specialising} the generic \corres
rules for each of these constructs to the particular \srel and types used in the
input program. This in effect discharges the non-trivial assumptions of these
rules once-and-for-all, allowing the automated proof of correspondence to 
proceed efficiently.

Conceptually, the refinement proof proceeds bottom-up, starting with the leaf functions of
the program and ending with the top-level entry points;  \corres results
proved earlier are used to discharge \corres assumptions for callees.
The \corres proof tactic thus follows
the call-graph of the input program. Currently, the tactic is limited to
computing call graphs correctly only for programs containing up to second-order
functions. We did not need higher orders in our applications yet, but
the tactic can certainly be extended if needed.

The resulting refinement theorem at this stage assumes that \corres holds for 
all the abstract functions used in the 
 program.
\begin{theorem}
Let~$f$ be a (monomorphic) \CDSL function, such that $\FunDefn{f} = \FunDef{f}{\tau \rightarrow \tau'}{x}{e}$. Let~$p_m$ be its monadic shallow embedding, as
derived from its generated C code. Let~$u$ and $v_m$ be arguments of
appropriate type for~$f$ and $p_m$ respectively. Then:
\[
\begin{array}{l}
\forall \mu\ \sigma. \;\valrel\ u \ v_m \longrightarrow \corres\ \srel\ e\ (p_m\ v_m)\ (x \mapsto u)\ (x : \tau)\ \mu \ \sigma
\end{array}
\]
\end{theorem}

\subsection{From Update to Value Semantics}

To complete this step, the compiler simply applies 
\autoref{thm:updvalrefinement}.

\subsection{From Monomorphic to Polymorphic Deep Embedding}\label{s:mono-correctness}
\label{s:mono}
Having made the transition to the value semantics, the proof now establishes
the correctness of the compiler's monomorphisation pass, moving upwards
in \autoref{fig:refinement} from
a monomorphic to a polymorphic deep embedding of the input program.

In this pass, the compiler generates an injective renaming function \rename that, for a polymorphic function 
name $f_p$ and types $\many{\tau}$, yields the specialised 
monomorphic function name $f_m$, mapping names downwards, from 
the polymorphic to the monomorphic level.
Just as we assume abstract functions are correctly implemented in C, we
also assume that their behaviour
remains consistent under
\rename. 

To establish correctness of monomorphisation, we essentially have an Isabelle
function that repeats the monomorphisation process on behalf of the \CDSL compiler, 
and prove that (1)~the monomorphised
program it produced is identical to that produced by the compiler, and
(2)~that the monomorphised program is a correct refinement of the polymorphic
one.
We define two Isabelle/HOL functions, both parameterised by~$\rename$: 
one for monomorphising expressions,
called $\monoexpr$, and the other for monomorphising (function) values,
called \monoval. 
The functions specialise function calls and use 
\rename to monomorphise 
all function calls in expressions and values, respectively.  
The functions are defined compositionally for all other \CDSL constructs. 

Step~(1) is proved by straightforward rewriting, and is automated on a per-program basis. 
Step~(2) is embodied in the following refinement theorem,
which we prove, once and for all,  by rule induction over the value semantics. 
The specialisation Lemma~\autoref{lemma:spec} of \autoref{sec:typing:poly}, 
is a key ingredient of this proof.

\begin{theorem}[Monomorphisation]
Let $f$ be a (polymorphic) \CDSL function whose definition given by
$\FunDefn{\cdot}$ is $f\ \VarN{x} = e$. Let~$v$ be an appropriately-typed
argument for~$f$. Let~$\rename$ be an injective renaming function. Then:
\[
\begin{array}{l}
\forall v'.\  \ValSem{(x \mapsto \monoval \ \rename \ v)}{\monoexpr\  \rename \ e}{\monoval\ \rename\ v'}
\longrightarrow \ValSem{(x \mapsto v)}{e}{v'}
\end{array}
\]
\end{theorem}

\noindent Note that on the left-hand-side of the implication, the computation
runs under the value semantics where the renaming is applied across
the $\FunDefn{\cdot}$ of the right-hand side.

The compiler generates a well-typedness proof for the monomorphic deeply
embedded program (\autoref{s:typetree}). We use the top-level theorem's injectivity assumption on
\rename to infer well-typedness of the polymorphic deeply embedded program.

\subsection{From Deep to Shallow Embedding}

\newcommand{\extfb}{\mathrm{ext2\_free\_branch}}
\newcommand{\valRel}{\mathrm{valRel}}

In this section, the proof makes the transition from deep to shallow
embedding, where the shallow embedding is a pure function in Isabelle/HOL. 
This shallow embedding is still in A-normal form and is produced
by the compiler as a separate Isabelle/HOL theory file. There is a second,
neater shallow embedding, explained in the following section, that is closer
to the \cdsl input program.

For each \CDSL type, the compiler generates a corresponding Isabelle/HOL type
definition, and for each \CDSL function, a corresponding Isabelle/HOL
constant definition. We can drop the linear types at this stage and remain in
Isabelle's simple types, because we have already made use of them: we are in
the value semantics.

In addition to these definitions, the compiler produces a theorem that the
deeply embedded polymorphic \CDSL term under the value semantics correctly
refines this Isabelle/HOL function. Refinement is formally defined here
by the predicate $\scorres$ that defines when a shallowly embedded
expression~$s$ is refined by a deeply embedded one~$e$ when evaluated
under the environment~$V$.

\begin{definition}[Deep to Shallow Correspondence]
$$\scorres\ s\ e\ V\ \equiv\ \forall r.\ \ValSem{V}{e}{r} \longrightarrow \valRel\ s\ r$$
\end{definition}

\noindent That is, $s$ corresponds to $e$ under variable bindings $V$ if whenever $e$
evaluates to an $r$ under $V$, then $s$ and $v$ are in the value relation
$\valRel$. Similarly to the proof from monadic C to update semantics, the
value relation here is one polymorphic constant in Isabelle/HOL, defined
incrementally via ad-hoc overloading.

The program-specific refinement theorem produced is:
\begin{theorem}[Deep to Shallow Refinement]
Let $f$ be an A-normal \CDSL function such that
$\FunDefn{f} = \FunDef{f}{\pi}{x}{e}$, and
let~$s$ be $f$'s shallow embedding. Then
$$\forall v_s\ v.\ \valRel\ v_s\ v \longrightarrow \scorres\ (s\ v_s)\ e\ (x \mapsto v)$$
\end{theorem}

\noindent
Note that $\valRel\ v_s\ v$ ensures that $v_s$ and $v$ are of matching type,
and that the shallow expression $s\ v_s$ ensures in Isabelle's type system
that it is the appropriate one. Like the C refinement proof in
\autoref{s:c-to-deep}, this proof is automatic and driven by a set of
syntax-directed \scorres rules, specialised to \CDSL A-normal form.

\subsection{From Shallow Embedding to Neat Shallow Embedding}

\autoref{fig:shallow} depicts the final top-level shallow embedding, only mildly polished for
presentation, for the \cdsl example of \autoref{fig:cdsl-snippet}.

\begin{figure}[ht]
\begin{lstlisting}[style=isa]
ext2_free_branch (Cnt.mk depth nd (Acc.mk ex fs inode) mdep) \<equiv>
if depth + 1 < mdep then 
case uarray_create (RR.mk ex (to\<^sub>f nd - fr\<^sub>f nd)) of
  R\<^sub>1\<^sub>1.Success ds\<^sub>1\<^sub>0 \<Rightarrow>
    let (ex, ds\<^sub>1\<^sub>2) = take ds\<^sub>1\<^sub>0 RR.p1\<^sub>f;
        (children, ds\<^sub>1\<^sub>3) = take ds\<^sub>1\<^sub>2 RR.p2\<^sub>f;
        (mbuf, nd_t) = take nd mbuf\<^sub>f;
        (children, ds\<^sub>1\<^sub>6) = take
          (uarray_map_no_break
            (ArrayMapP.mk children (fr\<^sub>f nd_t)
              (to\<^sub>f nd_t) ext2_free_branch_entry
              (Cnt.mk ex inode (fr\<^sub>f nd_t) mbuf) ()))
          RR.p1\<^sub>f;
  ...    
\end{lstlisting}
\caption{Shallow embedding for the example from \autoref{s:overview}.}
\label{fig:shallow}
\end{figure}

\noindent As \autoref{fig:shallow} shows, the Isabelle definitions use the same names
as the \cdsl input program and they have the same structure as the input
program. In this example, it remains visible that the compiler replaces
tuples from the surface syntax with records in the core language, e.g.\
\code{Cnt.mk} is the Isabelle record constructor for the type \code{Cnt}, and
instead of tuple pattern matching, the compiler generates a sequence of
\code{take} expressions. In practice, these disappear by rewriting when 
reasoning about the function. Tuple syntax could be reconstructed
in an additional small proof pass if so desired.

The correctness statement for this phase is simple: it is pure Isabelle/HOL
equality between the A-normal and neat shallow embedding for each function.
For instance:
$$\mathrm{Shallow.}\extfb = \mathrm{Neat.}\extfb$$
The proof is simple as well. Since we can now use equational reasoning with
Isabelle's powerful rewriter, we just unfold both sides, apply extensionality
and the proof is automatic given the right congruence rules and equality
theorems for functions lower in the call graph. This proof stage was the
easiest and fastest of the stages to construct; it took about 1 person day.

This is a strong indication that this representation of the program is well
suited for further reasoning on top.

\section{Discussion and Lessons Learned}\label{s:lessons}

\paragraph{Language Restrictions: Totality}
The current version of \CDSL purposefully omits primitive constructs for
iteration and recursion, because we wanted to ensure that the language was
total for a neat shallow embedding in HOL (which is total). However, since
our language meta-level proofs do not require totality, we only require that
\emph{each program} is terminating. We are therefore contemplating to relax
this restriction and allow \CDSL iterator constructs where termination is
obvious enough for Isabelle to prove automatically.

\paragraph{Formal Language Semantics}
The \CDSL semantics in Isabelle departs slightly from that presented in
\autoref{s:lang}. In particular, we enriched the update semantics to carry
enough value type information to infer their corresponding C types, and
adjusted the typing rules accordingly. While not needed for any of the
proofs of \autoref{s:lang}, this information is used in the automatic
C-correspondence proof.
In addition, we found ourselves repeating parts of the (linear) type
preservation proof in rule inductions on the semantics that make use of
typing assumptions. This means, while type erasure is an important
property for languages to enjoy (and \emph{is} enjoyed by \CDSL), dynamic
semantics with type information are helpful for mechanised
reasoning. Ideally, there should be an erased and a typed dynamic
semantics, with type safety implying their equivalence.

%

\paragraph{Optimisation}

The current \CDSL compiler performs little optimisation when generating C
code and leaves low-level optimisation to gcc or CompCert. Clever
optimisations in the \cdsl-to-C stage would complicate our current
syntax-directed correspondence approach. \CDSL-to-\CDSL optimisations,
however, are different. The ease by which we prove the correctness of the
A-normalisation over the shallow embedding via rewriting, suggests fruitful
ground for optimisation. We leave exploring this idea for future work.

\newcommand{\totalEffort}{$\approx 5$ person-years\xspace}

\newcommand{\compilerEffort}{$\approx 10$ person-months\xspace}

\newcommand{\certcompEffort}{$\approx 33.5$ person-months\xspace}

\newcommand{\langproofEffort}{$\approx 18$ person-months\xspace}

\newcommand{\isabelleLOC}{$\approx 17,000$ lines of code (including comments and whitespace)\xspace}

\newcommand{\compilerSLOC}{$\approx 9,500$ source lines of code (excluding comments and whitespace)\xspace}

\newcommand{\exttwoloc}{6,454\xspace}
\newcommand{\extproof}{76,759\xspace}

\paragraph{Effort and Size} \CDSL has been
under development for over 2 years and has continually evolved
as we have scaled the language to ever larger applications.
All up, the combined language development
and certifying compiler took \totalEffort. Engineering the
\CDSL compiler, excluding \certcompEffort spent on
proof automation and proof framework development,
consumed \compilerEffort. The remaining \langproofEffort was for the
design, formalisation and proof of \CDSL and its properties (e.g. the
theorems of \autoref{s:lang}), a small amount of which was also spent on
early compiler development.
The total size of the development in
the Isabelle theorem prover is \isabelleLOC, which includes the
once-and-for-all language proofs plus automated proof tactics to perform
the translation validation steps, given appropriate hints from the \CDSL
compiler. The \CDSL compiler, written in Haskell, is \compilerSLOC.
For \exttwoloc lines of etx2 \cdsl code we generate \extproof lines of Isabelle/HOL proofs
and embeddings.

\section{Related Work}\label{s:related}

Like us, the High-Assurance Systems Programming~\citet{Habit:lang} pro\-ject 
seeks to improve systems software by combining formal 
methods and programming language research. Like \CDSL, 
HASP's systems
language, Habit, is a domain specific functional language. 
\citet{McCreight_CT_10} show the correctness of a garbage collector in this project; however, to the best 
of our knowledge, there exist no full formal language semantics yet. 

Ivory~\citep{Pike_HBEDL_14} is a domain specific language embedded in Haskell
also for implementing correct systems software. It generates well-defined,
memory safe C code; however, unlike \CDSL it does not \emph{prove} correctness
of the generated code.

Linear types have been used in several general purpose imperative languages
to ensure memory safety without depending on a runtime, such as in Vault~\citep{Fahndrich_DeLine_02} and \citet{Rust:lang}.
PacLang \citep{Ennals_SM_04} is an imperative domain-specific language which uses linear
types to guide optimisation of packet processing applications on network
processors. Similar substructural type systems, namely uniqueness types, have been
integrated into functional programming languages such as
Clean~\citep{Barendsen_Smeters_93}. However, the type system there is only
used as a way to provide a purely functional abstraction over effects, 
and thus Clean still depends on a run-time garbage collector.


To the best of our knowledge, \citet{Hofmann_00} is the only work which proves the equivalence of the functional and imperative interpretation of a language with a linear type
system. The proof is by pen and paper, from a first order functional language with linear types to its translation in C. \CDSL in comparison is higher order and its compiler
produces a machine checked proof linking a purely functional shallow embedding
to its C implementation.


Examples for verified compilers for high-level languages are
CakeML~\citep{Kumar_MNO_14}, discussed in more detail in \autoref{s:intro}, which compiles a full ML dialect, including verified runtime and garbage
collection. In contrast, \citep{Neis_HKMDV_15} focuses on a compositional approach to
compiler verification for a relatively simple functional language, Pilsner,
to an idealised assembly language.

\citet{Chargueraud_10, Chargueraud_11} also generate a shallow
embedding representation of a program to facilitate proofs about
properties via a proof assistant, as we do. However, they do not address the
verification of the code generated by the compiler.

\section{Conclusions}\label{s:concl}

We have presented the \cdsl language, its self-certifying compiler, their
formal definitions and top-level compiler certificate theorem,
and the correctness theorems for each compiler stage. The
language targets systems code where data sharing is minimal or can be abstracted,
performance and small memory footprint are requirements, and formal
verification is the aim.

\cdsl is a pure, total functional language to enable productive equational
reasoning in an interactive theorem prover. It is higher-order and
polymorphic to increase conciseness. It uses linear types to make memory
management bugs compile time errors, and to enable efficient destructive
in-place update. It avoids garbage collection and a trusted runtime to reduce
footprint. It supports a formally modelled foreign-function interface to
interoperate with C code and to implement additional data types, iterators
and operations.

It does all of these with full formal proof of compilation correctness
and type-safety in Isabelle/HOL.

\CDSL sets a new benchmark for trustworthy systems languages, and
demonstrates, through the careful application of language design
with verified compilation in mind, that writing systems code that
supports purely functional equational reasoning is possible.

%

\balance
{
  \Finalfalse

  \ifAlpha
    \bibliographystyle{alpha}
  \else
    \bibliographystyle{plainnat}
  \fi
  \ifFinal
    \errmessage{Replace this by the content of your .bbl file!}
  \else
    \bibliography{references}
  \fi
}
\end{document}